\documentclass{aa} 
\usepackage[varg]{txfonts}
\usepackage{amsmath,mathtools}
\usepackage{graphicx}
\usepackage[T1]{fontenc}
\usepackage{comment}
\usepackage{orcidlink}
\usepackage{enumitem}
\bibpunct{(}{)}{;}{a}{}{,} 

\hypersetup{breaklinks=true,colorlinks=true, citecolor=blue, linkcolor=blue,urlcolor=blue}

\newcommand{\Msun}{\mathrm{M}_{\odot}}
\newcommand{\hi}{\ifmmode{\rm HI}\else{H\/{\scshape i}}\fi}
\newcommand{\msunyr}{{\rm M}_\odot \, {\rm yr^{-1}}} 
 
\newcommand{\kms} {\,{\rm km\,s}^{-1}}
\defcitealias{Tress+25}{T25}

\title{Simulating winds in the Galactic centre}
\subtitle{ I. Supernova-driven multiphase outflows and \hi\ cloud acceleration}

\titlerunning{Supernova-driven multiphase outflows in the Galactic centre}

\author{Nicolas~Peschken\inst{1,2}\thanks{Contact e-mail: \href{mailto:nicolas.peschken@unifi.it}{nicolas.peschken@unifi.it}}\orcidlink{0000-0002-9925-2974}
 \and  Enrico~M.~Di Teodoro\inst{1,2}\orcidlink{0000-0003-4019-0673}
 \and  Michał~Hanasz\inst{3}\orcidlink{0000-0002-2370-5631}
 \and  Lucia~Armillotta\inst{1,2}\orcidlink{0000-0002-5708-1927}
 \and  \\ Robin~Tress\inst{4}\orcidlink{0000-0002-9483-7164}
 \and  Thorsten~Naab\inst{5}\orcidlink{0000-0002-7314-2558}
 \and  Dominik~W\'{o}lta\'{n}ski\inst{3}
 \and  Artur~Gawryszczak\inst{6}\orcidlink{0000-0002-4435-4594}
 }
\institute{Department of Physics and Astronomy, University of Florence, Largo E. Fermi 2, I-50125 Firenze, Italy
\and INAF - Osservatorio Astrofisico di Arcetri, Largo E. Fermi 5, I-50125 Firenze, Italy
\and Institute of Astronomy, Nicolaus Copernicus University, ul. Grudziadzka 5/7, PL-87-100 Toruń, Poland
\and École Polytechnique Fédérale de Lausanne, observatoire de Sauverny, Chemin Pegasi 51, 1290 Versoix, Switzerland 
\and Max Planck Institute for Astrophysics, Karl-Schwarzschild-Str. 1, 85748 Garching, Germany
\and Nicolaus Copernicus Astronomical Centre, Bartycka 18, 00-716 Warsaw, Poland
}

\begin{document}

\abstract{The centre of the Milky Way (MW) hosts powerful multiphase outflows, as evidenced by the Fermi and eROSITA bubbles, and by cold atomic hydrogen (\hi) gas clouds detected up to a few kiloparsecs above the disc. In this paper, we investigate the process of launching gaseous outflows in the nuclear region of our Galaxy from supernova feedback. Using the {\scshape Piernik} code, we perform a simulation of the Galaxy with 3~pc resolution in both the Central Molecular Zone (CMZ) and the surrounding outflows. Our model follows the entire gas dynamics, from accretion onto the central star-forming ring through the dust lanes to star formation, feedback and the launching of outflows. Star formation occurs in cycles of starbursts followed by quiescent periods, mainly driven by intermittent gas inflows along the dust lanes. Stellar feedback generates hot ($\sim$10$^7$ K) winds launched from the CMZ at velocities of order $1000~\kms$, as well as colder ($\sim$10$^4$ K) \hi\ gas clouds with velocities of $\sim$100 $\kms$ at heights of $1-2$ kpc from the mid-plane. The spatial distribution, kinematics, and masses of our simulated clouds are broadly consistent with observations. Their properties indicate that they are accelerated out of the disc by entrainment from the hot phase. At least $20\%$ of these clouds return to the disc in fountain flows, while the majority are disrupted by interaction with the hot phase. While periods of intense star formation and supernova activity lead to more numerous outflowing clouds with higher masses and densities, quiescent phases with star formation rates close to that observed in the CMZ still produce \hi\ clouds consistent with data. 
These results suggest that stellar feedback alone, operating in a time-variable nuclear environment, can account for the observed population of cold clouds in the Galactic centre outflow.
}

\keywords{Galaxy: centre – Galaxy: kinematics and dynamics - Magnetohydrodynamics (MHD) - Methods: numerical}

\maketitle

\section{Introduction}

The nuclear region of the Milky Way (MW) is an extreme environment, hosting intense star formation, a central supermassive black hole (Sgr A$^*$), and large-scale gaseous outflows. Evidence for such outflows comes from a variety of multi-wavelength observations, highlighting the presence of high-velocity multiphase gas above and below the Galactic centre (GC). In particular, the Fermi bubbles, extending up to $\sim$10 kpc in $\gamma$-rays \citep{Su+10}, together with the eROSITA bubbles observed in soft X-rays \citep{Predehl+2020}, provide compelling evidence of fast outflows containing both hot thermal gas with temperatures $T>10^6$ K and non-thermal particles. On smaller scales, the discovery of high-velocity atomic hydrogen (\hi) clouds ($T\simeq10^4$ K) up to at least $\sim$2 kpc from the Galactic plane \citep{McClure-Griffiths+13,DiT+18}, some of which contain cold ($T\simeq10^{1-3}$ K) molecular cores \citep{DiTeodoro+2020,DiTeodoro+2026,Heyer+2025}, also suggests the action of powerful winds capable of accelerating dense gas to high speeds.
These features are further supported by observations of high-velocity warm ionised gas ($T\sim10^5$ K) detected in ultra-violet absorption within the Fermi bubbles \citep{Fox+15,Bordoloi+17,Cashman+2023}. 

The driving mechanisms behind these nuclear outflows in the MW remain debated. Galactic winds are often attributed either to active galactic nuclei (AGN) or to elevated star formation. While Sgr A$^*$ is currently quiescent, past episodes of AGN-like activity could plausibly explain the observed outflows, in particular the formation of the Fermi bubbles \citep[e.g.,][]{Zubovas+2012,Bland-Hawthorn+2019}. Alternatively, winds could be powered by stellar feedback, in the form of stellar winds and supernova (SN) events, associated with intense star formation activity in the nuclear region \citep[e.g.,][]{Crocker+2015,Sarkar+2015}. 
In fact, the nucleus of the MW hosts a region of extreme star formation, referred to as the ``Central Molecular Zone'' \citep[CMZ, e.g.,][]{Morris1996, Henshaw2023}. The CMZ is characterized by a ring-like structure of dense molecular gas of radius $\sim$100--300 pc formed from gas funnelled inward by bar-driven torques \citep[e.g.,][]{1992MNRAS.259..345A,Mazzuca2008}. The physical conditions at the CMZ are extreme: high turbulence \citep{2017ApJ...839....1L,Sormani+19}, frequent cloud collisions \citep{2019MNRAS.488.4663S}, and gas densities ($\sim$10$^{4-5}$ cm$^{-3}$ in molecular clouds) larger than in the rest of the Galaxy \citep{2016MNRAS.457.4536W}. Despite its high gas content and extreme densities, its present-day star formation rate (SFR) is relatively modest, $\sim$0.08 $\Msun$ yr$^{-1}$ \citep{2013MNRAS.429..987L,Henshaw2023,2024ApJ...962...14H}, and has been relatively constant over the past $\sim$5~Myr \citep{2017MNRAS.469.2263B}. The dynamics and evolution of the central regions of the MW have been extensively studied with hydrodynamical simulations \citep[e.g.,][]{Sormani+15,Sormani+17,Sormani+20,Shin+2017,2019MNRAS.490.4401A,Armillotta+2020,Tress+20,Tress+25}, providing valuable insights into the interplay between gas inflows, star formation, feedback, and instabilities. However, these efforts have primarily focused on the CMZ itself. Because the spatial resolution in such simulations is usually density-dependent, the diffuse outflowing gas is under-resolved, making it difficult to follow the formation and evolution of both the hot wind and possible cold clouds within it. 

Instead, SN-driven multiphase outflows have been investigated in detail thanks to dedicated high-resolution simulations, such as stratified-box (also called ``tall-box'') simulations \citep[e.g.,][]{2017ApJ...846..133K, 2020ApJ...900...61K, 2023MNRAS.522.1843R, 2025MNRAS.539.1706V}.
By focusing on a small patch of a galaxy, these simulations can resolve both star formation and outflows down to a few parsecs, capturing at once the structure of the interstellar medium (ISM) and the development of winds up to several kiloparsecs from the disc.
In addition, the interactions between different gas phases in outflows have also been extensively studied using wind-tunnel simulations, which follow the evolution of a single cold cloud travelling within a hot wind at parsec-scale resolution \citep[e.g.,][]{Scannapieco&Bruggen15,2016MNRAS.455.1309B,Gronke+19,Gronke+20,2021MNRAS.505.1083G}.  
All these works provided key insights on the creation and evolution of multiphase winds: while hot outflows naturally arise from the blowout of SN-driven superbubbles, the presence of cold gas can be explained either in terms of entrainment, i.e., clouds are lifted up from the disc and accelerated within the hot wind \citep[e.g.,][]{Schneider+18,Tan+24}, or ``in situ'' formation, i.e. clouds form directly within the hot wind via thermal instabilities \citep[e.g.,][]{2016MNRAS.455.1830T,Zhang+2018}.
Simulations also showed that the outflow mass loadings can be dominated by the cool phases, especially close the launching region, while the energy loadings are always dominated by the hot phase. The resulting loading factors are highly dependent on the environmental conditions in the disc, and in particular on the gas surface density and on whether SNe are clustered or distributed more uniformly. While some simulations also consider environments with star formation and gas surface densities similar to those found in the GC \citep[e.g.][]{2017ApJ...841..101L,2020ApJ...900...61K}, they do not model the gas dynamics and gravitational potential of the inner region of our Galaxy. This is crucial for reproducing the CMZ, which is constantly evolving due to gas inflows along the bar, the fast-rotating ring-like structure, and the formation of transient asymmetries \citep{Sormani+17}.

In this paper, we present a new simulation of the MW with parsec-scale resolution in the central regions, performed using the hydrodynamical code {\scshape Piernik} (\citealt{2010EAS....42..275H, 2010EAS....42..281H, 2012EAS....56..363H, 2012EAS....56..367H}). The simulation self-consistently follows radial gas accretion through the Galactic bar, the formation and evolution of a CMZ-like ring, star formation, supernova feedback, and the generation of gaseous outflows. Unlike previous simulations of the CMZ, our model simultaneously resolves the CMZ's ISM and the outflows it generates, enabling a full characterization of the launching and evolution of a multiphase wind. 
Our aim is to study the processes at play in creating nuclear outflows in our Galaxy, and, by comparing the simulation predictions to multi-wavelength observations of the MW’s nuclear winds, to better understand their physical origin and evolution. 
In this first paper of a series, we present our model and focus on characterizing SN-driven outflows from the CMZ, with a particular emphasis on the properties of cold gas clouds embedded in the wind.
After describing our {\scshape Piernik} simulation of the MW in Section \ref{presentation}, Section \ref{sec:gas_dynamics} will be dedicated to presenting and discussing the gas dynamics of the central regions and the launching of multiphase outflows from it. In Section \ref{sec:Clouds}, we will focus on the cool neutral phase, associated with \hi\ gas, analysing the properties of outflowing gas clouds in our simulation.
Finally, we discuss our results in Section~\ref{discussion} and conclude in Section \ref{ccl}.

\section{Presentation of the {\scshape Piernik} simulation}
\label{presentation}

\subsection{The {\scshape Piernik} code}
\label{PIERNIK}

Our simulation is run with {\scshape Piernik}, a grid-based multi-fluid magneto-hydrodynamical (MHD) code, using conservative numerical schemes\footnote{The public version of {\scshape Piernik} is available at 
\href{https://github.com/piernik-dev/piernik}{github.com/piernik-dev/piernik}.}. 
{\scshape Piernik} is parallelised on the basis of the Message Passing Interface (MPI) standard, and its data input/output communication utilises a parallel Hierarchical Data Format 5 (HDF5).
While the functionality of {\scshape Piernik} includes the modelling and mutual interactions of multiple fluids such as gas, dust, magnetic field and cosmic rays, the simulation presented in this paper includes only gas and magnetic field. 

The MHD algorithm is based on the standard set of ideal MHD equations:

\begin{align}
\begin{dcases}
  & \frac{\partial\rho}{\partial t} + \nabla\cdot\left(\rho\mathbf{v}\right) = 0 \\
  & \frac{\partial\rho\mathbf{v}}{\partial t} + \nabla\cdot\left(\rho\mathbf{v}\mathbf{v}^\mathrm{T} - \frac{\mathbf{B}\mathbf{B}^\mathrm{T}}{4\pi}\right) + \nabla P_\mathrm{tot} = \rho\mathbf{g}\\
  & \frac{\partial \varepsilon}{\partial t} + \nabla\cdot\left[\left(\varepsilon + P_\mathrm{tot}\right)\mathbf{v} - \frac{\mathbf{B}(\mathbf{B}\cdot\mathbf{v})}{4\pi}\right] = \rho\mathbf{v}\cdot\mathbf{g} + n\Gamma - n^2 \Lambda \\
  &\frac{\partial\mathbf{B}}{\partial t} - \nabla \times \left(\mathbf{v}\times\mathbf{B}\right) = 0 
  \end{dcases}
\end{align}

\noindent where $\rho$ and $n$ are the gas volume and number densities, $\mathbf{v}$ its velocity, $\mathbf{B}$ the magnetic field, $\mathbf{g}$ the gravitational acceleration, 
$\varepsilon$ the total energy density (kinetic + thermal + magnetic), $P_\mathrm{tot}$ the total pressure (thermal + magnetic), $\Gamma$ and $\Lambda$ the heating and cooling rate coefficients as detailed in Section~\ref{sec:therm}.
In {\scshape Piernik}, the ideal MHD equations are evolved with a finite-volume scheme using the Harten–Lax–van Leer Discontinuities (HLLD) Riemann solver \citep{2005JCoPh.208..315M}, together with the divergence-cleaning algorithm of \citet{2002jcoph.175..645d}, following the implementation described in \citet{2023MNRAS.522.5529P}.

{\scshape Piernik} includes a new dedicated Adaptive Mesh Refinement (AMR) module based on a block-based grid decomposition. Refinement is performed with a fixed factor of two and is restricted to create at most one level of difference within differential or interpolation stencil range. For more efficient use of computational resources, it allows for incomplete coverage of parent blocks, an approach that distinguishes it from, e.g., the PARallel Adaptive MESH framework \citep[PARAMESH,][]{Paramesh2000}. The load balancing assumes equal computational cost of each block at start, but it is possible to achieve more accurate adjustments based on profiling integrated into main solvers. The AMR module provides several options for the refinement prolongation order, as well as multiple refinement criteria. A basic option is the static refinement of nested regions with progressing levels of refinement. Other options include refinement criteria based on a threshold value, gradient, or relative gradient of any arbitrary variable (density, pressure, etc.).  Another criterion, applicable to self-gravitating systems, related to the Jeans criterion, requires that a given number of cells resolve the Jeans length. In simulations involving particles, the user can choose the criterion based on a maximum number of particles in a refinement block (which may allow for better load balancing and thus better scaling of massively parallel runs).  Additionally, users can define custom, problem-specific refinement criteria as needed.

\subsection{Cooling and heating terms}
\label{sec:therm}

Cooling is treated with an exact integration scheme \citep{2009ApJS..181..391T}, similarly to \cite{2023MNRAS.522.5529P}. The cooling rate coefficient, $\Lambda$, is computed from the gas temperature $T$ using a fitting formula by \cite{Koyama+02} for $T<10^4$~K, and a piece-wise power-law fit to the tabulated collisional ionization equilibrium cooling values at solar metallicity from \cite{1993ApJS...88..253S} for $T>10^4~$K.

Heating is treated using an explicit scheme.
We account for both photo-electric heating from the far-ultraviolet (FUV) radiation emitted by young massive stars and heating from ionizing cosmic rays, so that the total heating rate coefficient is $\Gamma = \Gamma_\mathrm{PE}+\Gamma_\mathrm{CR}$. To reflect the fact that intense star formation activity in the CMZ leads to a stronger interstellar radiation field and a higher rate of cosmic ray acceleration, we treat the CMZ region, defined as a cylinder of 400~pc radius and 100~pc height (the typical scale-height of the nuclear regions) and the rest of the galaxy differently. 

The photo-electric heating rate in the CMZ region is calculated as:

\begin{equation}
\Gamma_\mathrm{PE} = \Gamma_\mathrm{PE, 0} \, {e}^{-n/n_0}
\end{equation} 

\noindent where $\Gamma_\mathrm{PE, 0}$ is the photo-electric heating rate in the diffuse gas, and the exponential term represents attenuation due to local FUV shielding within dense clumps, with the corresponding turnover density $n_0 = 70$~cm$^{-3}$. The photo-electric heating rate in the diffuse gas is given by $\Gamma_\mathrm{PE, 0} = \Gamma_\mathrm{PE, \odot} \, J_\mathrm{FUV,0}/J_\mathrm{FUV, \odot}$, with $J_\mathrm{FUV,0}$ the typical FUV intensity in the CMZ region, and $\Gamma_\mathrm{PE, \odot}=2 \times 10^{-26}$~erg~s$^{-1}$ and $J_\mathrm{FUV,\odot}=2.1 \times 10^{-4}$~erg~s$^{-1}$~cm$^{-2}$~sr$^{-1}$ the photo-electric heating rate \citep{Koyama+02} and FUV intensity in the solar neighbourhood \citep{Draine78}, respectively. Based on simulations of CMZ-like environments by \citet{Moon_2021}, we adopt $J_\mathrm{FUV,0} = 10^{-2}$~erg~s$^{-1}$~cm$^{-2}$~sr$^{-1}$, which is representative of ambients with star formation and gas surface densities typical of the CMZ. The attenuation law and turnover density are also taken from \citet{Moon_2021}, who post-processed one of their simulations of CMZ-like environments with the adaptive ray-tracing algorithm developed by \citet{KimJ+17} to directly measure the FUV intensity produced by all star particles. Outside the CMZ region, the photo-electric heating rate is instead set to a constant value equal to $\Gamma_\mathrm{PE, \odot}$.

The cosmic ray heating rate in the CMZ region is set to a constant value, $\Gamma_\mathrm{CR} = 2 \times 10^{-25}$~erg s$^{-1}$, representative of environments with star formation and gas surface densities typical of the CMZ \citep{Moon_2021}. Outside the CMZ region, the cosmic ray heating is set to 0. As a result, the thermal equilibrium varies with height above the CMZ, as radiative cooling becomes increasingly dominant in the absence of cosmic-ray heating and due to the reduced photoelectric heating outside the disc. Moreover, both within and outside the CMZ, we assume $\Gamma=0$ for temperatures greater than $1.5 \times 10^4$~K, representing photo-electric and cosmic ray heating shutting off completely in fully ionised gas.

\subsection{Gravity}
\label{sec:gravpot}

{\scshape Piernik} includes a multigrid Poisson solver dedicated to problems involving self-gravity \citep{doi:10.1137/S1064827598346235}.  The solver is coupled with a multipole expansion method \citep{1977JCoPh..25...71J} to evaluate the gravitational potential outside the computational domain, which is then used to provide external boundary conditions for the multigrid solver within the domain. This ensures that the potential asymptotically follows a $1/r$ behaviour far from the domain centre and minimises unwanted effects due to the presence of finite computational domain boundaries.

A N-body particle-mesh solver for point masses representing stellar and dark matter components of galaxies was also introduced recently. The algorithm for N-body uses the leapfrog scheme together with particle mesh, projecting particle density to the grid using the Triangular Shaped Cloud (TSC) scheme. The resulting gravitational potential is then computed on the grid using the multigrid Poisson solver and the multipole solver. The TSC algorithm is used again to derive the particle accelerations. Particles outside the computational domain are treated with the multipole solver. In the simulation presented in this paper, N-body particles are used only for newly formed star clusters, as the dark matter halo and stellar disc are modelled via the external fixed potential described below.

To simulate the conditions of the MW's centre, and in particular of the CMZ, we implement an external, rotating gravitational potential based on \cite{2024A&A...692A.216H}. This potential represents a state-of-the-art description of the mass components in the inner Galaxy and has been optimised to satisfy several observational constraints, including recent measurements of the circular velocity \citep[e.g.,][]{Eilers+2019,Mroz+2019} and of the vertical acceleration \citep[e.g.,][]{Widmark+2022} in the MW.

\citeauthor{2024A&A...692A.216H}'s potential includes both axisymmetric and non-axisymmetric mass components. The axisymmetric components are a supermassive black hole, a nuclear star cluster (NSC), a nuclear stellar disc (NSD), a stellar disc (SD) and a dark matter halo (DM).
The non-axisymmetric parts of the potential include a stellar bar, modelled with three components representing its boxy-peanut shape and its ellipsoid short and long shapes, and gaseous spiral arms. 
For our simulation, we decided to neglect the black hole and spiral arm parts of the potential as they are not relevant for the scales investigated in this work: the former acts on scales much smaller than the CMZ (at radii $R\lesssim 1$ pc), while the latter is relevant only outside the bar region ($R\gtrsim5$ kpc).
Therefore, in our simulation, the density distribution of the stellar and dark matter components, from which the external gravitational potential is derived, is given by:

\begin{equation}
    \rho_\mathrm{tot} = \rho_\mathrm{NSC} + \rho_\mathrm{NSD} + \rho_\mathrm{bar,1} + \rho_\mathrm{bar,2} + \rho_\mathrm{bar,3} + \rho_\mathrm{SD} + \rho_\mathrm{DM} \quad .
\end{equation}

\noindent For brevity, we refer the reader to \cite{2024A&A...692A.216H} for a detailed description of each mass component and its parameters. In our simulation, the non-axisymmetric part of the potential, namely the bar, rotates with a pattern speed of $\Omega_\mathrm{bar} = 40$ $\kms$ kpc$^{-1}$ \citep[e.g.,][]{Li+2022,Clarke+2022}.

\subsection{Star formation and stellar feedback}
\label{sec:FB}

To model star formation, we use sink stellar particles accreting gas mass, similarly to \cite{2023MNRAS.522.5529P}. To form stars, a gas cell needs to fulfil the following conditions:  

\begin{enumerate}
    \item The gas temperature is lower than $10^4$~K. 
    \item The local velocity divergence is negative.  
    \item The cell's mass exceeds the Jeans mass.
    \item The gas dynamical time is greater than its cooling time.      
\end{enumerate}

\noindent When a gas cell meets all these conditions, it is considered star forming, and a fraction of its mass is accreted to a sink particle nearby (if no sink particle is found within a 10~pc radius, a new sink particle is created), following the SFR density:

\begin{equation}
    \rho_\mathrm{SFR} = \epsilon \sqrt{\frac{32 G\rho^3}{3\pi}}
\end{equation}

\noindent with $\epsilon$ being the star formation efficiency per free-fall time, set to 10\% \citep[e.g.][]{2013apj...770...25a,2018MNRAS.475.3511G}.
Once a sink particle reaches 100~$\Msun$, it is locked and is changed to a regular star particle, i.e.\ it cannot accrete mass any more. After 6.5~Myr 
(duration of stellar winds preceding SN events for massive stars, e.g. \citealt{2016ApJ...824...79A, Dubois_2021}), this particle undergoes a SN event, following the scheme described below. The choice to limit sink particles to a maximum mass of 100~$\Msun$ allows for short-lived sink particles, created and exploding within the CMZ, instead of long-lived sink particles coming from outside the CMZ, accreting mass in the CMZ and then exploding outside. This also avoids having very massive star particles clumping in the centre.  
    
Once a sink particle is changed to a star particle, and after the 6.5~Myr delay, the ejected gas energy corresponding to a single SN event ($10^{51}$~erg) is released at the current position of the star particle. To model this, we follow the scheme implemented by \citet{2017ApJ...846..133K} in their tall-box simulations of the ISM. 
We summarise the main features of the scheme here, but we refer the reader to \cite{2017ApJ...846..133K} for full details. The energy is deposited together with 10~$\Msun$ of gas isotropically within a sphere of 3 cell radius centred on the star particle position, corresponding to a physical radius of 9~pc in the enhanced-resolution CMZ region (Section~\ref{sec:IC}). We compute the enclosed gas mass within this sphere $M_\mathrm{SNR}$, as well as the expected shell formation mass $M_\mathrm{sf}=1679~\mathrm{M_{\odot}}(n_\mathrm{H}/\mathrm{cm^{-3}})^{-0.26}$, and use the ratio $R_{\mathrm M}=M_\mathrm{SNR}/M_\mathrm{sf}$ to determine whether the Sedov-Taylor phase of the supernova remnant is locally resolved or not. If $R_\mathrm{M}>1$, the Sedov-Taylor phase is unresolved, and the explosion is simply modelled by injecting momentum radially into the sphere, corresponding to the final momentum of the SN, $p_{\rm SN}=2.81\times10^5$~$\Msun$ $\kms$ $(n_\mathrm{H}/\mathrm{cm^{-3}})^{-0.17}$. If $0.027<R_\mathrm{M}<1$, the Sedov-Taylor phase is resolved, and we inject the SN energy of $10^{51}$~erg both in kinetic (28\%) and thermal (72\%) forms to the gas. In this case, to avoid extreme effects due to local under- or over-densities, we homogenise the densities, velocities and thermal energy to the mean values within the sphere, prior to the injection. If $R_\mathrm{M}<0.027$, the gas has very low density, and we inject the feedback as pure kinetic energy, while also resetting the densities, velocities and thermal energy to their mean values in the sphere. Contrary to \cite{2017ApJ...846..133K}, we keep a fix radius of 3 cells in the latter case, regardless of the enclosed mass. We systematically tested this feedback recipe by simulating single SN events in a uniform medium across a range of different ambient densities, and compared the resulting radial momentum and hot gas mass with those obtained from a pure thermal and a pure kinetic feedback models, similarly to \cite{2017ApJ...846..133K} who demonstrated that their fiducial model reproduces the expected outcome of a supernova remnant as predicted in higher-resolution simulations \citep{2015ApJ...802...99K}. We found results in very good agreement with \cite{2017ApJ...846..133K}, for the same resolution and at all densities.

\subsection{Initial conditions}
\label{sec:IC}

We simulate a MW-like galaxy within a computational domain of $50 \times 50 \times 25$ kpc, for a total volume of 62500~kpc$^3$.
The initial gaseous disc distribution is built in hydrostatic and thermal equilibrium with the axisymmetric part of the gravitational potential. The temperature in the mid-plane at a given cylindrical radius $R$ is assumed to be $T(R) = T_0\cosh \left( R/R_\mathrm{max} \right)$, with $T_0=6000$~K, and $R_\mathrm{max}=20$~kpc. This ensures a nearly uniform mid-plane temperature across the disc, with a smooth transition between the disc and the halo gas at $R=R_\mathrm{max}$. Furthermore, in the initial conditions, we include an ad-hoc heating term to stabilize the disk, making it thick enough to withstand the initial gravitational vertical collapse. The $z$-distribution (with $z$ being the vertical direction perpendicular to the disc plane) of temperatures is then iteratively derived from the gas momentum and energy equations, following \cite{2007MNRAS.376..861K}:
    
\begin{equation}
    \frac{dT}{dz} = \frac{m_{\mathrm H} g_{\mathrm z}}{k_\mathrm{B}(1+\alpha)} \frac{\Lambda(T)-G_0}{T\Lambda^\prime(T)-\Lambda(T)+G_0}  
\end{equation}

\noindent where $k_\mathrm{B}$ is the Boltzmann constant, $m_{\mathrm H}$ is the atomic mass of hydrogen, $g_{\mathrm z}$ is the $z$-derivative of the axisymmetric part of the gravitational potential, $\alpha$ the initial ratio between magnetic and gas thermal pressures, which we assume to be at equilibrium, $G_0=5 \times 10^{-26}$~erg~cm$^3$~s$^{-1}$ is the ad-hoc heating rate coefficient, assumed constant, $\Lambda$ is the cooling rate coefficient, and $\Lambda^\prime$ its derivative with respect to temperature. We compute temperatures starting from the mid-plane with $dz$ increments of 0.6~pc. 

The density distribution is then calculated from the temperature distribution using:

\begin{equation}
    n(T) = \frac{\Gamma_\mathrm{PE,\odot}}{\Lambda(T) - G_0} \quad 
\end{equation}

\noindent with $\Gamma_\mathrm{PE,\odot}$ the photo-electric heating rate at the solar neighbourhood (Section~\ref{sec:therm}).
Gas velocities are initialized to the circular speed determined by the axisymmetric part of the external gravitational potential.
The magnetic field is introduced in the Galactic disc with a scale-height of 100~pc, as a divergence-free toroidal field based on a vector potential, giving an initial strength of $\sim$0.1~$\mu$G in the CMZ region that is quickly amplified by several orders of magnitude due to turbulence and shearing inducing dynamo processes in the central region \citep[see e.g.][]{2022ApJ...924...26S}. 

To ensure the stability of the central parts of the Galaxy, we start without a bar in the gravitational potential, and gradually introduce the bar components linearly over one full bar rotation, i.e. 150~Myr \citep[e.g.,][]{1992MNRAS.259..345A,Tress+25}. 
After this phase, the bar is fully established, and its dynamical impact on the gas becomes evident, driving strong inflows toward the centre.
During the first 150~Myr, the spatial resolution is 6~pc in the CMZ, i.e. within the central 400~pc in the disc (up to $|z|=200$~pc). Outside the CMZ region, the resolution decreases gradually with increasing distance from the centre, reaching its lowest value of 390~pc at the periphery of the Galaxy. After 150~Myr, in order to better capture star formation and the resulting outflows, we increase the resolution to 3~pc in both the central region and in the outflows coming from it. Refinement is controlled by a combination of geometric criteria (up to 2~kpc height and 800~pc cylindrical radius) and density threshold (gas denser than 0.2~cm$^{-3}$). Note that each refined region has a size of at least (96~pc)$^3$ (corresponding to 32$^3$ cells), so that gas cells falling below the density threshold might still be resolved at 3~pc, as it is typically the case for cold gas clouds having a high peak density despite an average density lower than 0.2~cm$^{-3}$. The resolution elsewhere is left unchanged. The system is then evolved with this configuration for an additional 250~Myr, yielding a total simulation time of 400 Myr.
For our analysis, we only consider the high-resolution part of the simulation, i.e. $t>150$ Myr.

\section{Gas dynamics in the nuclear region}
\label{sec:gas_dynamics}
 
\begin{figure}
    \centering
     \includegraphics[width=0.5\textwidth]{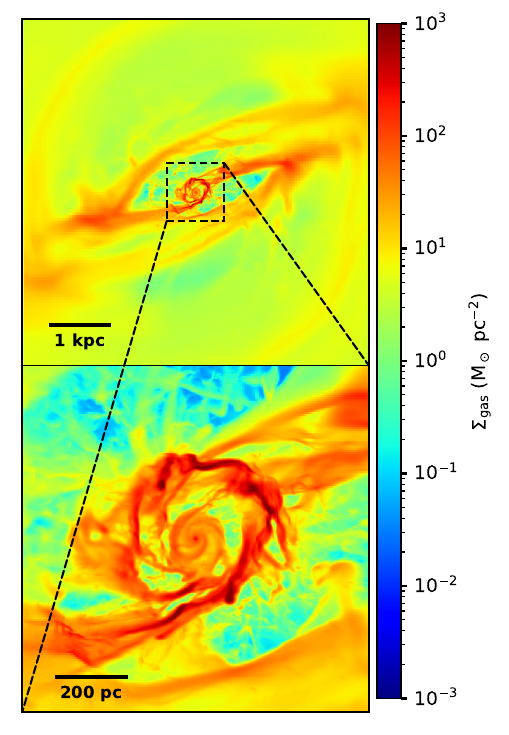}
     \caption{Projected gas density of our {\scshape Piernik}-simulated MW, with a zoom into the CMZ (bottom panel), at $t=160$~Myr, i.e. after full implementation of the bar component. The gas distribution shows a bar structure following the stellar component, of about 7~kpc length, which funnels gas towards the centre, fuelling the CMZ through two dust lanes. 
     }
     \label{fig:gal_CMZ}
\end{figure}

\begin{figure}
    \includegraphics[width=0.5\textwidth, trim={0.8cm 0 0.5cm 0},clip]{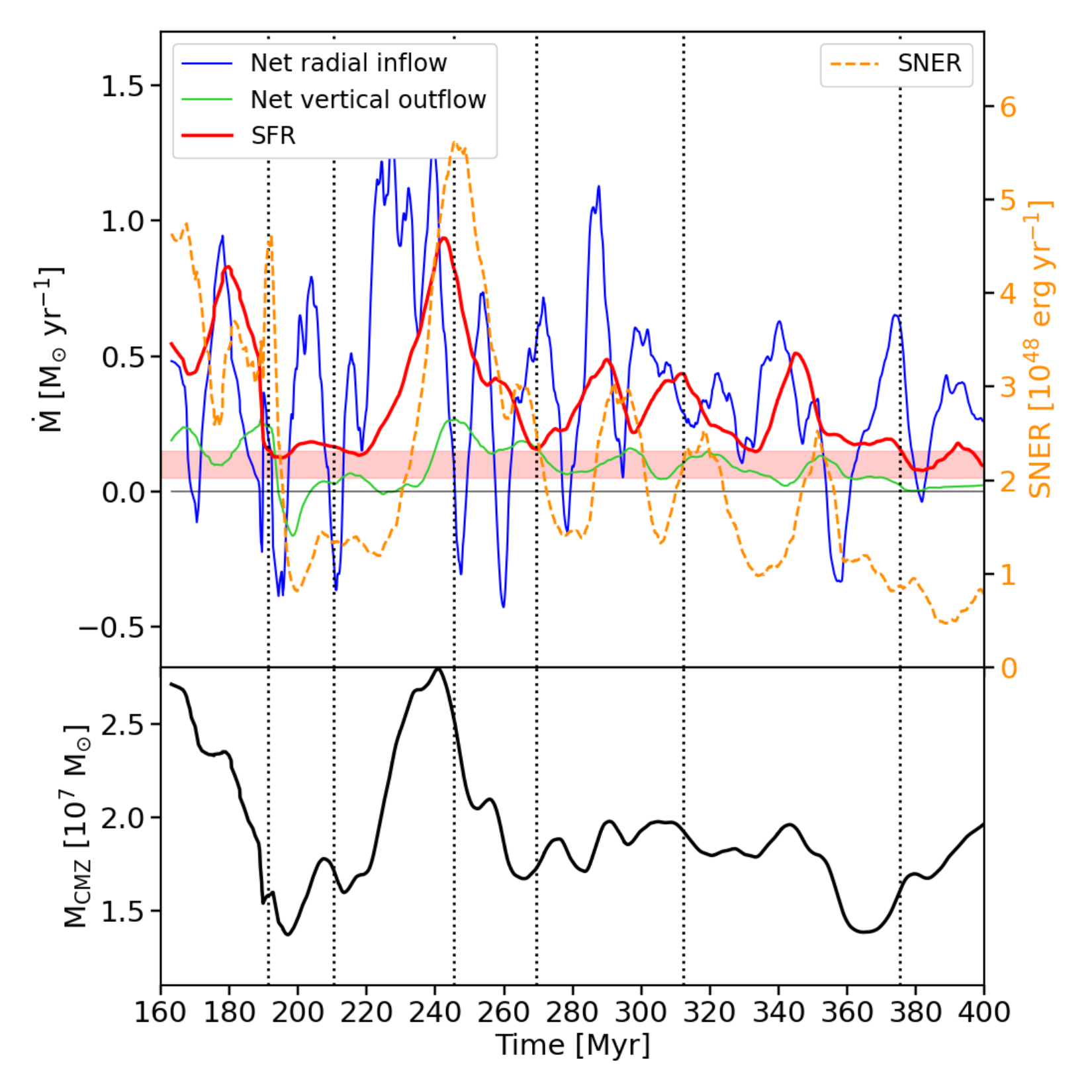}
    \caption{Top panel: Star formation rate (SFR, red), supernova energy rate (SNER, dashed orange), net radial mass inflow rate (blue), and net vertical mass outflow rate (green) at $|z|=300$~pc within the CMZ ($R<500$~pc), as a function of time, smoothed with a 10~Myr moving average. The observational range of values of the SFR in the present-day CMZ ($0.07^{+0.08}_{-0.02}~\msunyr$ from \citealt{Henshaw2023}) is shown as a red shaded band. 
    Bottom panel: total CMZ gas mass (also at $R<500$~pc) as a function of time. 
We indicate with vertical dotted lines six times of interest chosen based on the SFR and SNER.}
    \label{fig:sfr_sner}
\end{figure}

\begin{figure*}
    \centering
    \includegraphics[width=\textwidth]{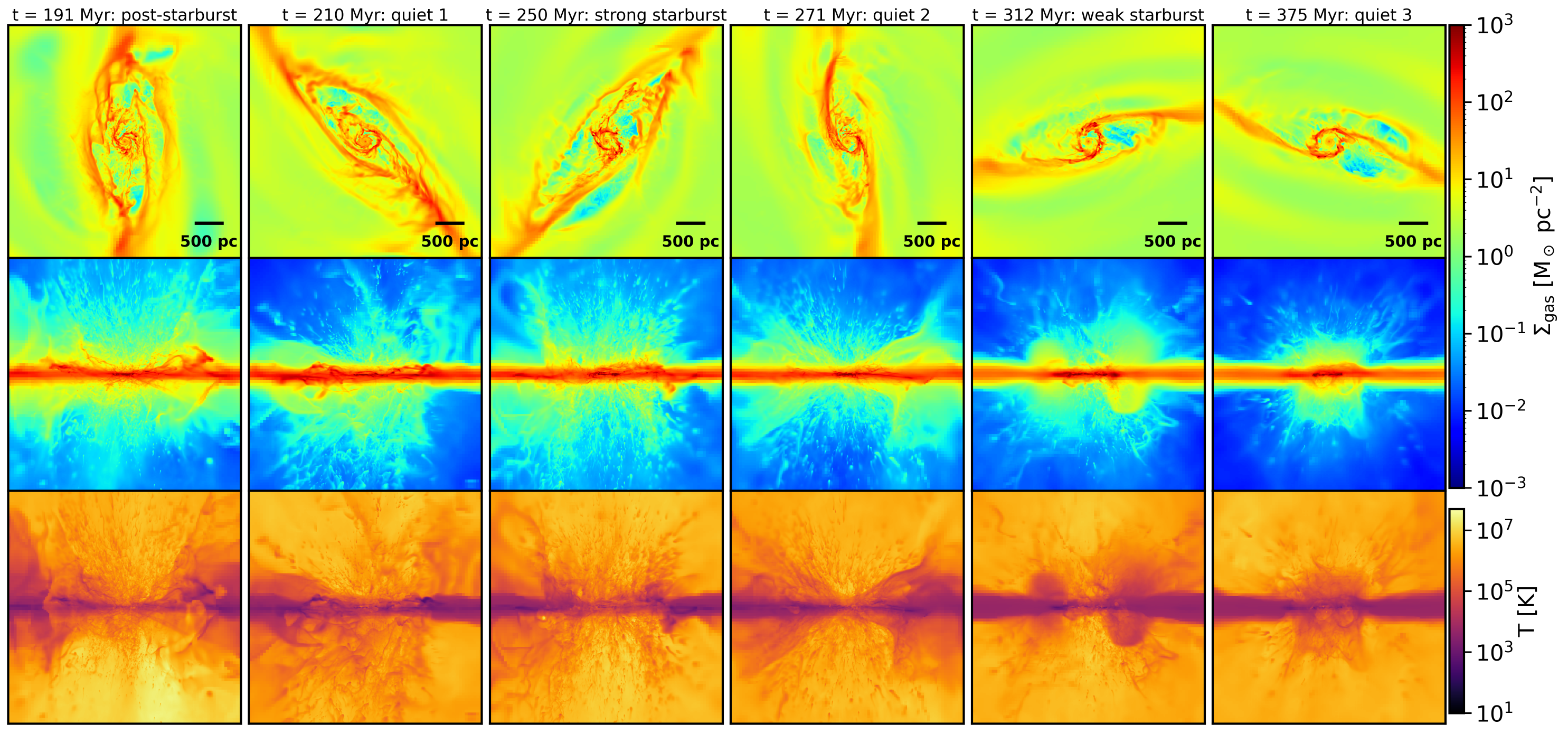}
    \caption{Projected gas density of the simulated GC seen face-on (top panels), edge-on (middle panels), as well as mass-averaged temperature seen edge-on (bottom panels), for the representative times highlighted in Fig.~\ref{fig:sfr_sner}. From left to right: post-starburst, first quiet star formation phase, strong starburst, second quiet phase, weak starburst, and third quiet period. The projection and averages are calculated over a 4~kpc box around the GC. }
    \label{fig:display_times}
\end{figure*}

\subsection{Central Molecular Zone and star formation cycles}
\label{sec:CMZ}
   
In our simulation, a ring-like gaseous structure, analogous to that in the MW's CMZ, naturally forms under the action of the external gravitational field. 
The morphology and properties of this structure are set primarily by the bar potential in the central Galaxy. The formation of the ring can be understood in terms of gas responding to the $x_1$ and $x_2$ orbital families that arise in barred potentials \citep{1991MNRAS.252..210B}. While the $x_1$ orbits are elongated parallel to the bar major axis, the $x_2$ orbits are smaller and elongated perpendicular to it. Gas flows inward along streamlines associated with progressively more elongated $x_1$-like orbits, until these become self-interacting. At this point, gas is violently shocked and loses angular momentum \citep{2015MNRAS.449.2421S,2019MNRAS.488.4663S} streaming towards the centre along the so-called bar’s dust lanes \citep{1992MNRAS.259..345A} and accumulating into a dense, ring-like structure associated with $x_2$ orbits. 
Fig. \ref{fig:gal_CMZ} shows the gas column density at time $t=160$ Myr in the inner 5 kpc (top panel) of our simulated Galaxy, together with a zoom-in of the central 1 kpc (bottom panel). A well-defined gaseous ring with radius $\sim$200~pc and gas surface density of $\sim$100~$\Msun$ pc$^{-2}$ has formed. Two bar-driven dust lanes connect to the ring, continuously feeding the CMZ with fresh gas inflowing from larger radii. This ring shows persistent asymmetries throughout most of the simulation, in agreement with observations of the MW centre \citep[e.g.][]{2025ApJ...984..156B} and with previous simulations showing that transient asymmetries can arise spontaneously from hydrodynamical and thermal instabilities in the GC \citep{2018MNRAS.475.2383S}.
As gas accumulates in the CMZ, it becomes gravitationally unstable, triggering star formation and stellar feedback that are predominantly concentrated within the ring.
Supernova events inject mass, momentum, and energy into the ISM, launching gaseous outflows out of the disc plane\footnote{A video showing the time evolution of the CMZ's surface density is available on the {\scshape Piernik} website, \href{https://piernik.umk.pl/results/2026a}{piernik.umk.pl/results/2026a}.}.

The evolution of the CMZ is therefore driven by the interplay between gas accretion, star formation, and outflows. 
We investigate the relation between these processes by calculating the time evolution of the SFR and of the supernova energy rate (SNER), i.e.\ the energy injected by SNe per unit time.
To quantify the effect of stellar feedback in removing gas from the CMZ, we also define a net mass outflow rate as the amount of gas mass crossing a horizontal surface per unit time, which we calculate as $\dot{M}_\mathrm{out, net}=\sum_i \rho_i v_{z,i} S_{i}$, with $\rho_i$, $v_{z,i}$ and $S_i$ being respectively the volume density, vertical velocity, and surface of each $i$-th cell in the layer immediately below the chosen horizontal plane.
This definition yields a net rate, i.e.\ it represents the difference between outflowing and inflowing gas across the surface.
Analogously, we quantity the effect of gas accretion from the dust lanes by using a radial mass inflow rate, i.e. the net rate of gas mass crossing the lateral surface of a cylinder of a given radius and height.
In the top panel of Fig. \ref{fig:sfr_sner}, we show the time evolution of the SFR (red line), together with the SNER (orange dashed line), the net vertical outflow rate (green line) and the net radial inflow rate (blue line).
In this plot, the SFR and the SNER are calculated within a region of $R<500$ pc, i.e.\ a radius about twice the size of the star-forming ring.
The vertical outflow rate is taken at a height $|z|=300$~pc, i.e. about three times the typical scale height of the inner disc region, while the radial inflow rate is calculated at $R=500$ pc and up to a height of $|z|=300$~pc.

Fig.~\ref{fig:sfr_sner} highlights that, in our simulation, star formation proceeds through cycles of starburst and quiescent phases on timescales of a few tens of Myr.
The SFR varies significantly over time, with values ranging from $\sim$0.1 $\msunyr$ to almost 1 $\msunyr$ in the CMZ, with mean and median values of 0.34~$\msunyr$ and 0.30~$\msunyr$, respectively. 
The corresponding SFR surface density in the ring has typical values of the order of a few~$\Msun$~yr$^{-1}$~kpc$^{-2}$, but it also fluctuates strongly throughout the simulation.
The current observed SFR in the CMZ, i.e.\ $0.07^{+0.08}_{-0.02}~\msunyr$ (\citealt{Henshaw2023}, red shaded band in Fig.~\ref{fig:sfr_sner}), is roughly consistent with the lowest values found in our simulation (e.g. at 200 Myr and 370 Myr). 
We note that while the SFR in the CMZ as inferred observationally from gas-based tracers appears relatively steady over timescales of a few Myr \citep[e.g.,][]{2017MNRAS.469.2263B}, stellar population studies also indicate that on longer timescales, the MW centre, and particularly the nuclear stellar disc, has experienced significant variability in its star formation activity. For example, \cite{2020NatAs...4..377N} estimates an average SFR over the past $30$ Myr in the nuclear stellar disc of $0.2-0.8~\msunyr$, a range of values that overlaps with the SFR oscillations in our simulation. 

The SNER varies in the range $1-5\times10^{48}$ erg yr$^{-1}$ and follows the oscillations of the SFR with a delay of approximately 5~Myr, consistent with the 6.5~Myr delay introduced in our stellar feedback scheme (see Section \ref{sec:FB}). $\dot{M}_\mathrm{out,net}$ closely follows the SNER, 
showing the role of feedback in pushing gas out of the disc: more SN events drive out larger quantities of gas. We note that, during periods of low SN activity (e.g., $200-230$ Myr), the net outflow rate becomes negative, indicating that more gas at $|z|=300$ pc is moving downward than upward. 
Over the full simulation, $\dot{M}_\mathrm{out,net}$ oscillates between $\pm0.3$ $\msunyr$.

\begin{figure*}
    \includegraphics[width=\textwidth]{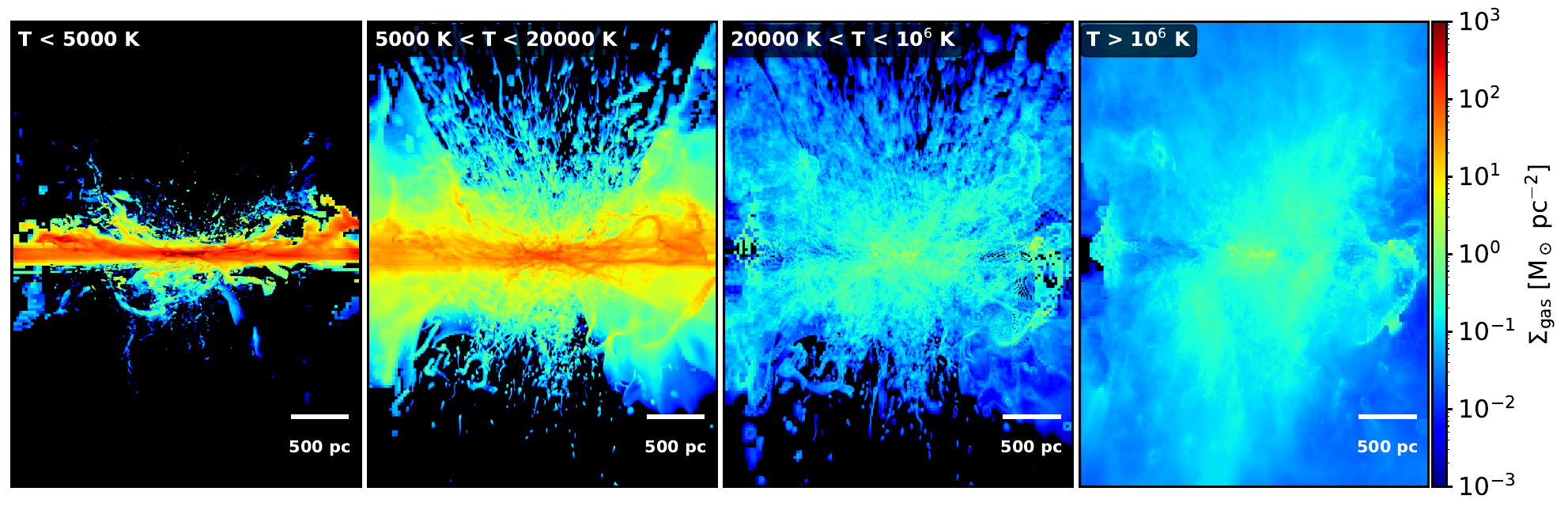}
    \caption{Edge-on projected density for gas in different temperature bins, at $t=191$~Myr (post-starburst period). From left to right: cold ($T<5000$ K), cool ($5000 < T < 2 \times 10^4$ K), warm ($2 \times 10^4 < T < 10^6$ K) and hot gas ($T>10^6$ K). }
    \label{fig:display_phases}
\end{figure*}

Temporal variations in the SFR (and thus feedback) reflect corresponding variations in the available gas mass.
While radial inflows supply fresh gas to the CMZ, star formation and outflows remove mass from it. 
In the bottom panel of Fig. \ref{fig:sfr_sner}, we show the mass of gas within the CMZ as a function of time.
As expected, the SFR closely tracks the amount of gas available in the CMZ, which in our simulation tends to be on the lower end of observational estimates \citep[$2-6 \times 10^7~\Msun$, e.g.,][]{Dahmen+1998,Ferriere+2007}. 
In turn, the CMZ gas mass also responds to the net radial accretion along the dust lanes, rapidly growing during episodes of strong inflow (e.g., around 240 Myr).
The radial inflow rate shows large fluctuations on short timescales (few Myr), reflecting the highly dynamical nature of the accretion process. Its values range from $1~\msunyr$ to $-0.5~\msunyr$, with negative values indicating periods during which more gas radially leaves the CMZ than enters it.
At several epochs, the inflow rate is consistent with observational estimates of $0.8\pm0.6~\msunyr$ \citep{2021ApJ...922...79H}, although these constraints remain highly uncertain.
Star formation and feedback also regulate the available gas mass, with episodes of intense star formation generally followed by a decline in the CMZ mass (e.g.\ around 180 Myr), and thus in SFR. 
In particular, SN feedback acts through a dual mechanism: it removes gas via outflows and simultaneously suppresses radial accretion from the dust lanes. We verified this with a control simulation without stellar feedback, which significantly alters the CMZ's evolution and morphology. Without SN-driven disruptions, the CMZ becomes more stable and uniform, fed by a steady, laminar gas inflow through the dust lanes that continuously builds up its mass. Stellar feedback thus regulates accretion: during starburst phases, SNe in both the CMZ and the dust lanes inhibit accretion, deplete the gas reservoir, and suppress star formation; once star formation declines, accretion resumes, replenishing the reservoir and initiating a new cycle.

In summary, our simulation produces a central region with a dense, dynamical gaseous ring similar to the nuclear parts of the MW. The CMZ operates in a self-regulated regime, where periods of enhanced accretion trigger elevated star formation and feedback, which in turn reduce the gas mass and inflow, leading to subsequent quiescent phases. In the remainder of this paper, we focus on the SN-driven outflows launched from the CMZ, investigating their properties and their connection to the underlying star-forming environment.

\subsection{Multiphase outflows}
\label{sec:Outflows}

\begin{figure}
    \includegraphics[width=0.5\textwidth, trim = 0.5cm 0 0 0, clip]{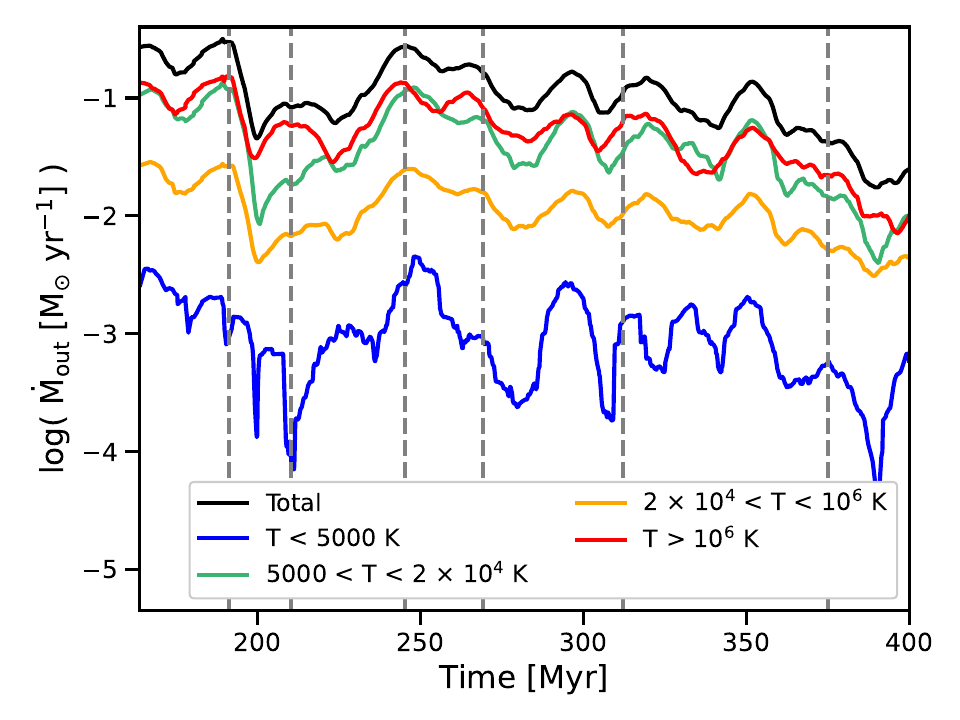}
    \caption{Mass outflow rate ($\dot{M}_\mathrm{out}$) above the GC as a function of time (moving average over 10~Myr) for the four temperature phases defined in the legend. $\dot{M}_\mathrm{out}$ is calculated at $|z| = 300$ and $R<500$~pc. The dashed vertical lines represent the six representative epochs shown in Fig. \ref{fig:display_times}.}
    \label{outf_time}
\end{figure}

\begin{figure}
    \includegraphics[width=0.5\textwidth,trim = 0.5cm 0 1.5cm 2.5cm, clip]{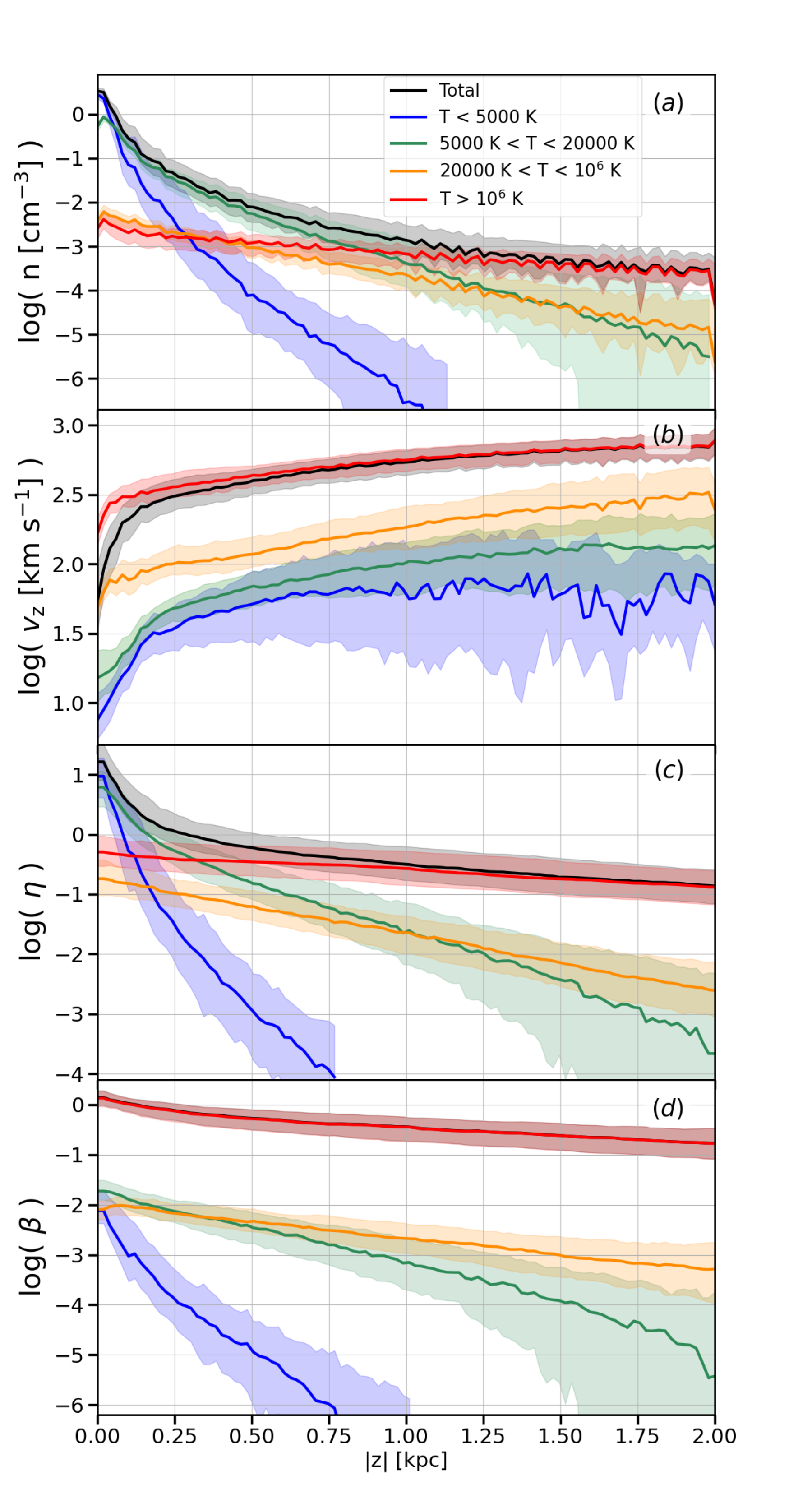}
    \caption{Vertical profiles for gas divided in four temperature phases (see legend in top panel), within $R=1$~kpc. Solid lines indicate median values over $t>180$~Myr, while shaded areas are between the 16$^{\rm th}$ and 84$^{\rm th}$ percentiles. From top to bottom: gas number density (panel $a$), vertical velocity $(b)$, mass loading factor (mass outflow rate divided by SFR) $(c)$, and outflow energy loading factor (energy outflow rate divided by SNER) $(d)$. For $|z| \gtrsim 1$~kpc, more than half of the snapshots contain no cold gas (blue curve), causing the median density, mass loading, and energy loading of this phase to drop to zero. Nevertheless, cold gas is present in some snapshots up to heights of $\sim$2~kpc, allowing the vertical velocity profile of the cold phase to extend beyond $|z| \gtrsim 1$~kpc.}
    \label{fig:zprofs_piernik}
\end{figure}

\begin{figure}
    \includegraphics[width=0.5\textwidth,trim = 0 0 0 0.4cm, clip]{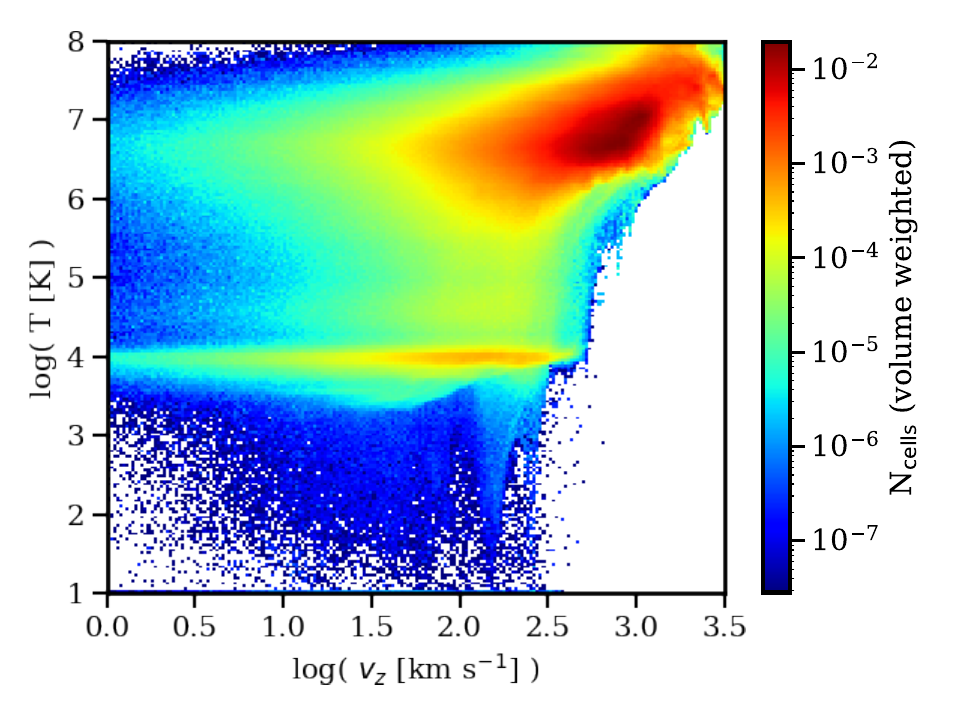}
    \caption{Temperature - vertical velocity diagram of gas with $0.3 < |z| < 2$~kpc during the post-starburst period, within $R=1$~kpc, weighted by volume. The phase diagram is dominated by two phases: a fast hot phase around 10$^7$ K, and a slow colder phase around 10$^4$~K.}
    \label{fig:vzT_diag}
\end{figure}

In the previous section, we discussed how the simulated CMZ evolves through different phases of star formation (Fig. \ref{fig:sfr_sner}). An initial strong star-forming phase, which reaches a ${\rm SFR}\simeq0.8~\msunyr$ around $t\sim180$~Myr, is followed by a quiescent period lasting until $t\sim220$~Myr, during which the SFR drops to values comparable to those observed in the present-day CMZ of our MW. We then have again another strong starburst around $t\sim240$~Myr followed by a lower SFR phase around $t\sim270$~Myr. After this, the CMZ undergoes a series of weaker starburst cycles, reaching peak SFRs of at most $\simeq0.5~\msunyr$. Near the end of the simulation, the CMZ returns to a quiescent state, matching once again the observed SFR value. 

In Fig.~\ref{fig:display_times}, we show the projected density (top and middle panels) and temperature (bottom panels) averaged along the line of sight, at six representative times of this star-forming cycle, highlighted as dotted vertical lines in Fig.~\ref{fig:sfr_sner}. From left to right, we plot $t=191$~Myr (transition from the initial starburst to the quiet period, but with still high SNER: post-starburst time), $t=210$~Myr (quiet star formation), $t=245$~Myr (strong starburst), $t=269$~Myr (quiet time 2), $t=312$~Myr (weak starburst), and $t=375$~Myr (quiet time 3). The edge-on projections of the gas density and temperature (second and third row, respectively) qualitatively show how the outflows vary across these different epochs. During starbursts, enhanced SN activity drives large amounts of warm and cool gas ($T\sim10^{4-5}$ K) up to 2~kpc and beyond, mostly in the form of small-scale ($\sim$10~pc) clouds. In quiescent periods, outflowing clouds are still present, but remain confined to regions closer to the disc, reflecting the lower SN activity. An interesting time is the post-starburst $t=191$~Myr, which comes about $10$ Myr after a strong starburst event. At this time, the SFR dropped to $\sim0.1$~$\Msun$ yr$^{-1}$, i.e. comparable to that of the present-day CMZ, but the SNER is still high, because of the time delay between star formation and SN events (Section \ref{sec:CMZ}). Therefore, the resulting outflows appear more prominent compared, for example, to the quiet period 15 Myr later, when the SFR is similar but the SNER has dropped by an order of magnitude. Towards the end of the simulation, the SFR becomes consistently quieter, and the SNER drops to very low values. This results in weaker outflows, making the third quiet time the least effective at driving cool winds, even though we still see some cool gas leaving the disc. During these late phases, the CMZ becomes more regular, and radial accretion along the dust lanes proceeds in a less chaotic manner, as clearly illustrated by the face-on density maps (top row of Fig.~\ref{fig:display_times}).

 \begin{figure*}
    \centering
    \includegraphics[width=\textwidth]{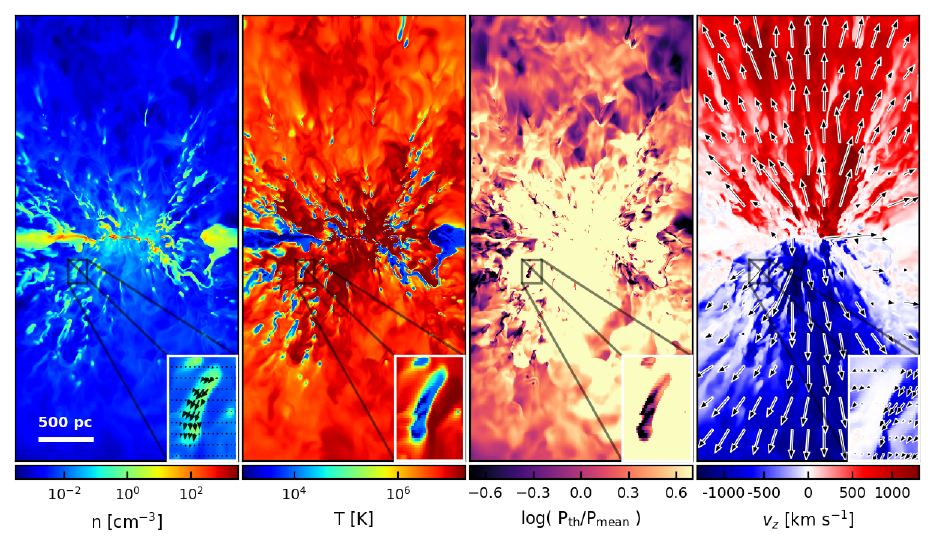}
    \caption{Slice plots of our simulation during post-starburst period. From the left to the right: gas number density, temperature, normalised pressure, and vertical velocity. For each panel, a zoom-in onto one cold cloud is shown in the bottom right corner. The arrows in the the first panel show the magnetic field direction, while in the fourth panel they show the gas velocity field. These plots clearly show the presence of dense, cool clouds outside the disc, with lower pressures and velocities than the hot phase, and a magnetic field typically aligned with the wind direction.}
    \label{fig:slices_plot}
\end{figure*}

To study the winds in more detail, we decompose the outflowing gas in four temperature phases: cold ($T<5000$~K), cool ($5000<T<2\times10^4$~K), warm ($2\times10^4<T<10^6$~K), and hot ($T>10^6$~K). To visualize the morphology of the different gas phases, in Fig. \ref{fig:display_phases} we show the projected gas density divided in these temperature bins for the post-starburst time ($t=191$~Myr).
While the hot phase (fourth panel) is very diffuse and volume-filling, the warm (third panel) and cool phases (second panel) are clearly present in the form of gas clouds and filaments in the region above and below the GC, extending up to heights of at least 2~kpc. 
Since \hi\ gas is typically associated with the cool phase, with characteristic temperatures of $T\sim10^4$~K, we therefore expect our simulation to produce a population of outflowing \hi\ clouds, similar to those observed in the central regions of the MW.
Colder and denser gas (first panel) is also found in the extraplanar region, although closer to the disc plane $(|z|< 1$~kpc), which may hint at the existence of a molecular component in the outflows. However, because our simulation does not include a proper chemical network, we cannot conclude on the presence of molecular gas in our model.

Fig. \ref{outf_time} illustrates the mass outflow rate ($\dot{M}_\mathrm{out}$) as a function of time for the four temperature phases, measured at $|z|=300$~pc above the CMZ  and  within $R<500$~pc. Unlike the net mass outflow rate $\dot{M}_\mathrm{out,net}$ discussed in Section~\ref{sec:CMZ} and plotted in Fig.~\ref{fig:sfr_sner}, we define $\dot{M}_\mathrm{out}$ considering only outflowing material, i.e. by summing over only gas cells that satisfy the condition $z\,v_z > 0$, with $z$ and $v_z$ being the height of the cell and the vertical component of the velocity, respectively. 
Knowing from Fig. \ref{fig:sfr_sner} that $\dot{M}_\mathrm{out,net}$ follows the SNER, and is therefore directly linked to the SFR with a delay of a few Myr, it is unsurprising that in Fig. \ref{outf_time} $\dot{M}_\mathrm{out}$ for the different temperature phases follows the same star-formation cycles. In particular, in periods of high SN activity, the mass outflow rate is large, reaching up total values of  $\dot{M}_\mathrm{out}\sim0.2-0.3~\msunyr$, while it decreases for all phases during quieter epochs. As mentioned, towards the end of the simulation the mass outflow rate tends to drop to much lower values due to decreased star formation and SN activity. At all times, at 300~pc from the mid-plane, the outflows are dominated by the hot and cool phases, with the warm gas being a factor of a few below, and the cold gas being about two orders of magnitude lower.

To quantify how gas at different temperatures evolve with height over the disc, in Fig.~\ref{fig:zprofs_piernik} we show median vertical profiles of gas number density $n$ (panel $a$), vertical outflowing velocity $v_z$ $(b)$, mass loading factor $\eta = \dot{M}_\mathrm{out} \, /\, \mathrm{SFR}$ $(c)$, and energy outflow loading factor $\beta = \dot{E}_\mathrm{out} \, /\, \mathrm{SNER}$ $(d)$, where $\dot{E}_\mathrm{out}$ is the energy outflow rate. At a given time and for each $|z|$ bin, we compute the gas densities simply as the bin's mass divided by its volume, while the velocities are averaged solely over the volume of outflowing cells in that given phase.
The energy outflow rate is defined, similarly to the mass outflow rate, as $\dot{E}_\mathrm{out} = \sum_i \varepsilon_i v_{z,i} S_{i}$, where $\varepsilon_i$ is the energy density (thermal + kinetic, magnetic energy being negligible in the outflows) of a given cell and the sum is taken only over outflowing cells in the layer immediately below the chosen horizontal plane.
We take the medians of these quantities over the entire simulation at $t>180$~Myr, i.e. after allowing 30 Myr for the system to adjust following the increase in resolution to 3 pc at $t=150$~Myr.
Hereafter, we focus on analysing the simulation only from $t=180$~Myr.
Moreover, since the outflows tend to be bi-conical and at large heights extend beyond the $R=500$~pc region considered so far for the CMZ, here we expand the analysis region to $R=1$~kpc.

The density of all gas phases (Fig. \ref{fig:zprofs_piernik}$a$) decreases with height more gradually in the hotter phases and more steeply in the colder phases.
The inner 100~pc from the mid-plane are dominated by cold gas (blue line) with $T<5000$~K, while cool gas (green) with $5000<T<2\times10^4$~K dominates the thick disc up to $|z|\sim800$~pc. 
Above this height, the hot phase ($T>10^6$~K, red) becomes the primary component, with the cool phase contributing only a small fraction of the total density at $|z|>1.5$ kpc.
All gas phases show a sharp acceleration in the inner $100-200$ pc, and their vertical velocity typically increases up to $|z|\sim2$ kpc, as shown in Fig.~\ref{fig:zprofs_piernik}$b$.
The hot gas vertical velocity tends to median values close to $\sim1000~\kms$; during strong outflowing periods, some hot gas can reach peak velocities of $\sim5000~\kms$.
The warm and cool phases move significantly slower than the hot phase, reaching median vertical velocities of $\sim300~\kms$ and $\sim100~\kms$ at $|z|\sim2$ kpc, respectively.
This dependence between temperature and velocity is also highlighted in Fig. \ref{fig:vzT_diag}, which shows the volume-weighted cell counts in the outflows ($0.3 < |z| < 2$~kpc) with a given temperature and vertical velocity at the post-starburst time.
Outflows are dominated by a fast-moving hot phase, with $T \sim 10^{6-7}$~K and $v_z \sim1000~\kms$, but a distinct, slower phase, with $T\sim10^4$~K and $v_z \sim100~\kms$, also clearly emerges.

Panel $(c)$ of Fig. \ref{fig:zprofs_piernik} shows that
the total mass loading factor (black line) is close to unity just above the disc, and decreases towards 0.1 at higher heights. 
The hot gas shows a mass loading factor that only weakly declines with height, from $\eta_\mathrm{hot} \simeq 0.4$ just above the disc to $\eta_\mathrm{hot} \simeq 0.1$ at $|z| = 2$~kpc, while the loading factors of the cold and cool phases also peak near the mid-plane ($\eta_\mathrm{cold} \simeq \eta_\mathrm{cool} \simeq 5$) but then sharply drop with height ($\eta_\mathrm{cold} = 0$ and $\eta_\mathrm{cool} \simeq 10^{-4}$ at 2 kpc). 
Thus, while the hot gas dominates the outflows at $|z|\gtrsim 300$ pc in terms of mass, the cool and cold phases carry most of the mass closer to the disc. When we look at the energy loading factor $\beta$, as shown in Fig. \ref{fig:zprofs_piernik}$d$, we find that the energy is almost entirely carried by the hot winds at all heights. For the hot gas,
$\beta_\mathrm{hot}\approx 1$ close to the disc and only slightly decreases with height, to $\beta_\mathrm{hot}\approx 0.1$ at $|z|=2$~kpc. The energy loading factors of the warm and cool gas are at least two orders of magnitude lower than that of the hot phase at all heights.

\begin{figure*}
    \includegraphics[width=0.33\textwidth,trim = 0.2cm 0 0.1cm 0, clip]{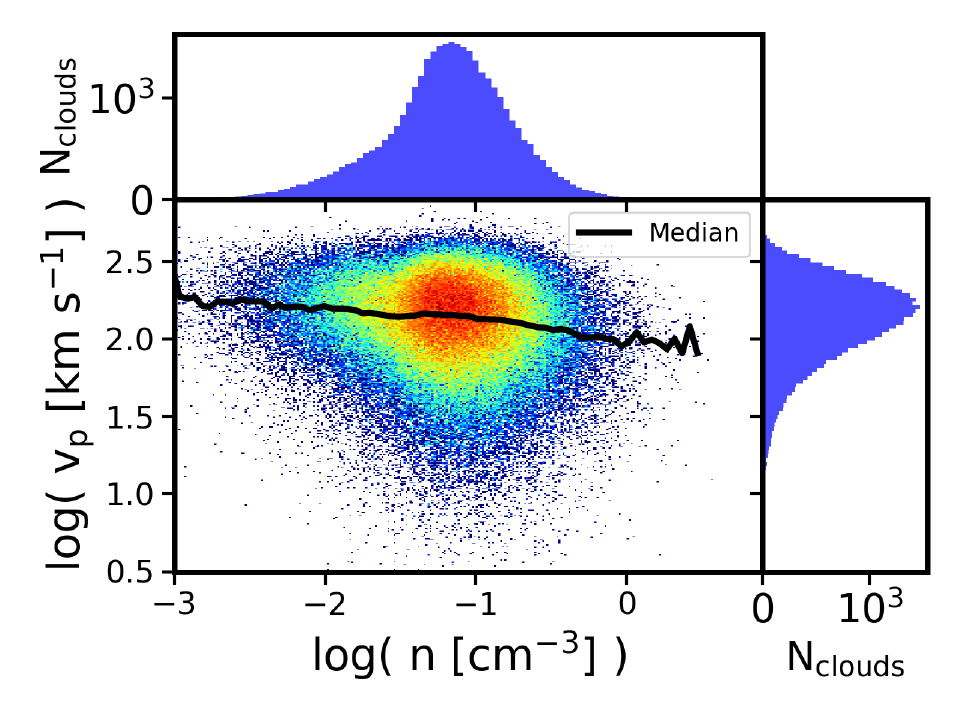}
    \includegraphics[width=0.33\textwidth,trim = 0.6cm 0 0.1cm 0, clip]{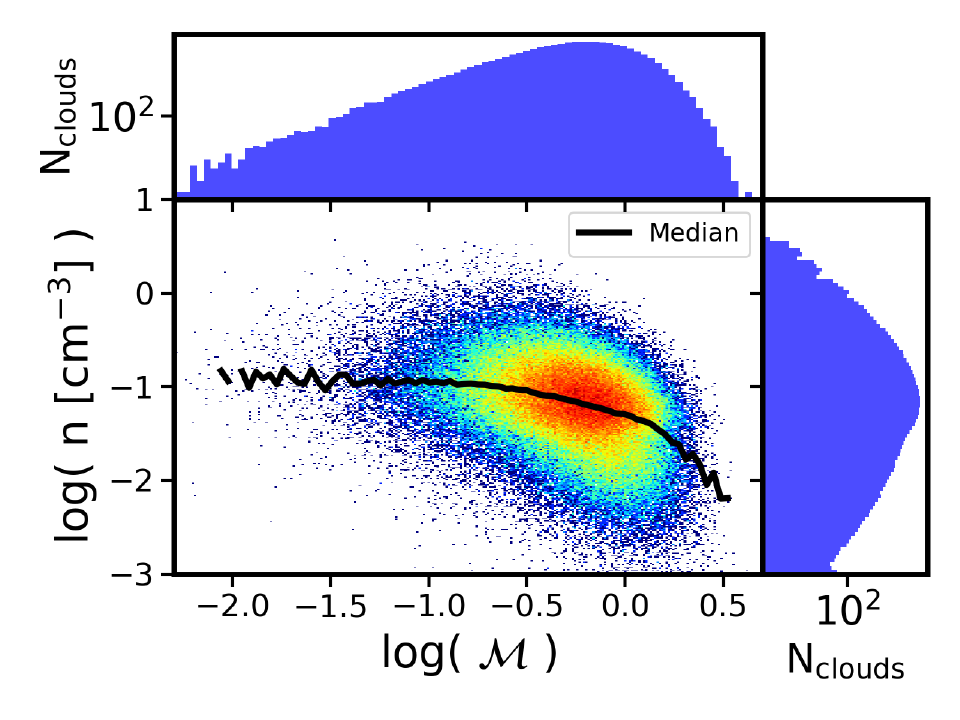}
    \includegraphics[width=0.33\textwidth,trim = 0.3cm 0 0.1cm 0, clip]{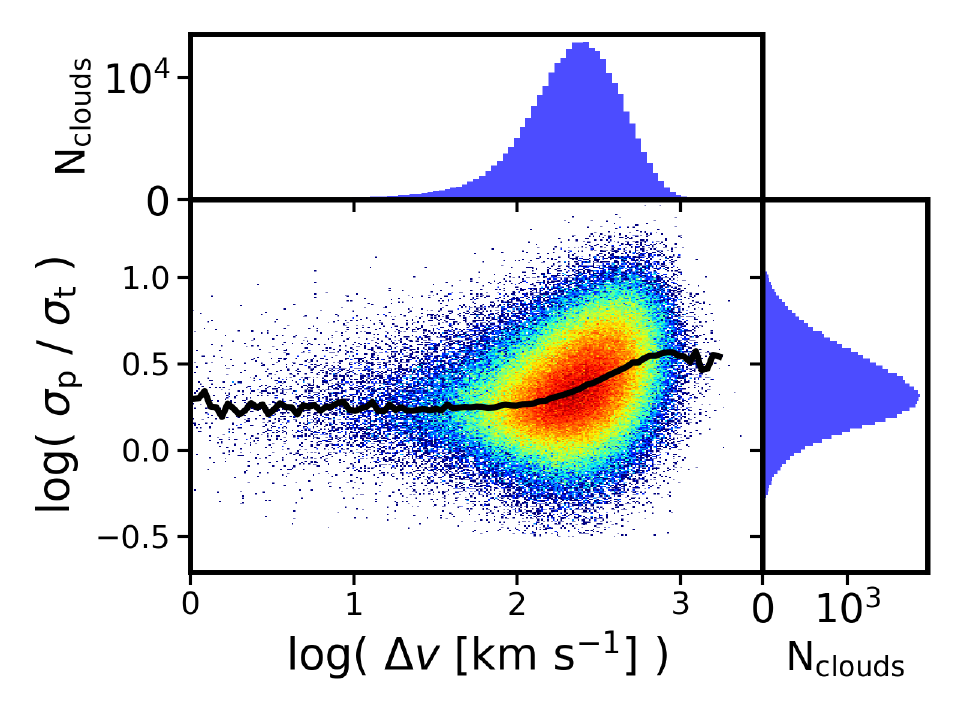}
    \caption{General properties of outflowing \hi\ clouds in the simulation. Left panel: poloidal velocity as a function of density, with the median shown in black. 
    Middle panel: cloud densities as a function of their Mach number (poloidal velocity of the cloud divided by the sound speed of the surrounding hot phase).
    Right panel: poloidal elongation of the clouds (defined as the ratio of poloidal to toroidal spreads $\sigma_\mathrm{p}/\sigma_\mathrm{t}$, see Section~\ref{sec:clouds_general}) as a function of velocity difference between the clouds and the surrounding hot wind (measured around each cloud individually).}
    \label{fig:clouds_dens_vel}
\end{figure*}

\section{\hi\ clouds in the outflows}
\label{sec:Clouds}

Fig.~\ref{fig:slices_plot} shows edge-on slice plots in the $x=0$ plane of gas density, temperature, thermal pressure and vertical velocity at the post-starburst time. The leftmost panel clearly shows the presence of over-densities in the outflows corresponding to cool clouds around 10$^4$~K (second panel). These clouds are also visible in the vertical velocity map (last panel), as they are outflowing but at much lower velocities than the surrounding hot gas, creating a trail behind them. Their temperature is characteristic of \hi\ gas, suggesting that these structures represent the simulated counterparts of the \hi\ outflowing clouds observed in the MW centre \citep[e.g.,][]{McClure-Griffiths+13,DiT+18}. We further show a zoom-in on one of these clouds in the bottom-right corner of each panel, with the first inset suggesting a magnetic field (arrows) preferentially aligned with the wind direction.
In this section, we focus on analysing the properties of these extra-planar \hi\ features in our simulation. 

To allow for a comparison to the \hi\ clouds observed above and below the GC, we approximate \hi\ gas in our simulation with the following approach. We assume that all gas with $T<10^4$~K is neutral hydrogen, while for gas with $T>10^4$~K we derive the \hi\ fraction as $X_\hi = 1 - X_{\mathrm{H}^+}$, where $X_{\mathrm{H}^+}$ is the fraction of ionised hydrogen at a given temperature. Assuming that hydrogen is fully ionised before the first helium ionisation occurs, we estimate the ionised hydrogen fraction as $X_{\mathrm{H}^+} = \mathrm{max}(X_{\mathrm i}-X_{\mathrm M},1)$, with $X_\mathrm{M} = 1.68\times10^{-4}$ being the fraction of metals with ionisation potential less than that of hydrogen \citep{2011piim.book.....D}. $X_{\mathrm i}$ is the total ionised fraction, which we approximate as $X_{\mathrm i}=\mathrm{min}(X_{\rm e},1.099)$, where the electron fraction $X_{\rm e}$ tabulated values are taken from \cite{1993ApJS...88..253S} for solar metallicity.

Our model produces a large number of \hi\ clouds and filamentary structures extending in the extra-planar region, up to at least 2~kpc from the CMZ, and following a bi-conical flow from the central region (see also Figs.~\ref{fig:display_phases} and \ref{fig:slices_plot}). 
To isolate these objects, we first construct a 3~pc resolution fixed-grid position-position-position (PPP) data cube of the \hi\ gas density, corresponding to the three Cartesian directions $x$, $y$, $z$, capturing a volume of $2 \times 2 \times 4$~kpc$^3$ centred on the GC.
We then apply the 3D source finder implemented in the $^{\rm 3D}$\textsc{Barolo} code \citep{2015MNRAS.451.3021D}, which identifies and reconstructs objects within a 3D dataset. Although this source finder is designed to operate with position-position-velocity (PPV) data cubes, we repurpose the third positional axis in our PPP cube as a pseudo-velocity axis to make the algorithm applicable. To avoid problems in regions containing no \hi\ gas, we add a small density background, and set this value as density threshold for the detection in the source finder. 

Wind tunnel simulations have shown that the evolution of cold cloud properties in interaction with a hot wind converges only at sufficiently high resolution, typically requiring approximately eight resolution elements per cloud radius \citep{Gronke+19}, or four in the subsonic regime \citep{Leary+2026}. As will be shown in the following section, most of our clouds are in the subsonic or transonic regime.
We therefore remove all clouds with less than 4 cells per clouds radius in the spherical approximation, which corresponds to 12~pc at 3~pc resolution (or $\sim$268 cells.)
Moreover, to mimic source-finding criteria used in observations \citep[e.g.,][]{DiT+18}, we exclude large filaments and structures by imposing a maximum size of 300~pc for sources in each direction. 
Finally, we remove sources detected within the disc, keeping only clouds with heights greater than 100~pc (corresponding to the typical scale-height of the CMZ region). 

The source finder was run for different simulation snapshots to isolate clouds in the outflows at different times.  The number of clouds detected this way varies between 30 during very quiet periods at the end of the simulation, and 1143 during periods of starburst, with a median value of 487.
Once we have identified a cloud, we trace its constituent cells back to the original simulation snapshot and extract or calculate some of their key thermodynamical properties (e.g., density, temperature, pressure) and kinematical properties (velocity and velocity dispersion). 
In addition to the cool cloud gas, we also analyse the properties of the surrounding warmer gas to characterise the ambient medium in which the clouds reside and with which they interact.
We stress that our current simulation does not allow us to follow easily the evolution of individual clouds within the wind. In future work, we plan to implement gas tracers to track clouds down and investigate their properties along their full orbits. In this paper, we focus instead on a statistical analysis of the cloud properties in order to investigate the main mechanisms at play in their formation and evolution.

\subsection{General properties}
\label{sec:clouds_general}

To get reliable statistics, we cumulate clouds found throughout the simulation (an analysis of the fluctuations of cloud properties with time and SNER will be presented in Section \ref{sec:clouds_time}), by extracting clouds every 0.5~Myr, for a total of 216178 clouds. Note that due to this small time interval, a given cloud is likely to be sampled several times across its evolution, so that the actual number of clouds present in the outflows throughout the simulation is necessarily lower. We find that 81.0\% of these clouds are outflowing (i.e., they have $z\,v_z>0$), while the remaining 19\% are falling back onto the disc (inflowing, $z\,v_z<0$). In this section, we focus only on the outflowing ones, and therefore exclude the inflowing clouds from the analysis.

Because the global outflow and the cloud trajectories are not perfectly vertical, but rather bi-conical (see Fig. \ref{fig:slices_plot}), the simple vertical velocity component $v_z$ would underestimate the actual outflowing speed of the clouds. 
To avoid this, we calculate instead the poloidal component of the velocity $v_\mathrm{p}$, derived as:

\begin{equation}
v_{\mathrm p} = \sqrt{v_\mathrm R^2+v_\mathrm z^2} = \sqrt{\frac{(xv_\mathrm x+yv_\mathrm y)^2}{x^2+y^2} +v_\mathrm z^2}
\end{equation}

\noindent where $v_\mathrm R$, $v_\mathrm x$ and $v_\mathrm y$ are the velocity components in the radial, $x$ and $y$ directions, respectively. The poloidal velocity allows us to capture radial trajectories, without taking into account the disc's rotation (toroidal component). Hereafter, we will use the poloidal definition when referring to the cloud velocities.

In the left panel of Fig. \ref{fig:clouds_dens_vel}, we show density -- poloidal velocity 2D histograms of the detected clouds, together with their 1D marginalised distributions. For a given cloud, its velocity is derived with a mass-weighted average over all cells composing it, while its density is simply its total mass (sum of cell masses) divided by its volume (sum of cell volumes).  
Most \hi\ clouds have densities spanning from $10^{-2}$ cm$^{-3}$ to 1 cm$^{-3}$, with the most common value being around $10^{-1}$ cm$^{-3}$.
The velocity distribution peaks around $v_\mathrm{p}\sim150~\kms$, but in the high-speed tail, clouds can reach outflowing velocities up to $\sim500~\kms$. 
Moreover, we note that higher density clouds tend to have lower outflowing velocities than low density ones, as also highlighted by the black thick line denoting the median velocity as a function of density. This behaviour can be attributed either to the greater inertia of denser clouds, which makes them harder to accelerate, or to the fact that faster clouds have experienced more disruptive interactions with the surrounding hot phase.

Similarly, in the middle panel of Fig. \ref{fig:clouds_dens_vel}, we display the joint distributions of cloud density and Mach number $\mathcal{M}$, defined as the ratio of the poloidal velocity to the sound speed of the surrounding hot gas. Hereafter, the surrounding hot phase (or hot wind) for each cloud is taken as all the gas with temperature greater than 10$^6$~K within a box three times the size of the cloud. Most clouds are sub-sonic (84.4\%), and those with larger Mach numbers (close to or exceeding unity) tend to be less dense. This decreasing trend is less clear when cloud density is plotted as a function of velocity, suggesting that the Mach number is a more informative indicator of cloud density evolution and potentially pointing to mass loss as clouds become supersonic.

\begin{figure}
    \includegraphics[width=0.5\textwidth]{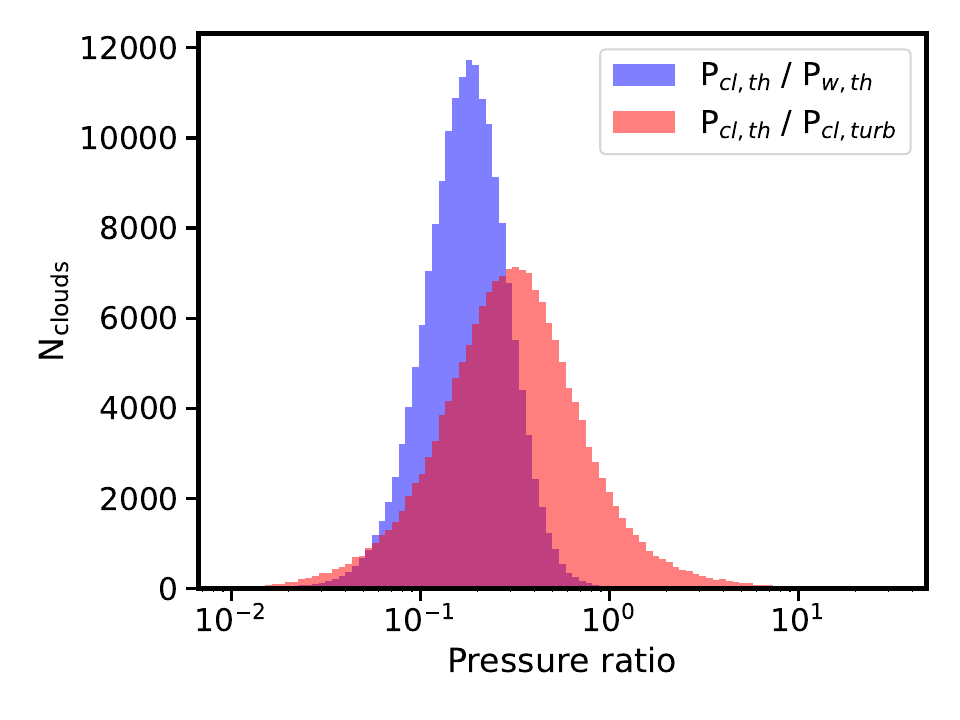}
    \caption{Pressure ratio distributions of all clouds: cloud thermal pressure divided by the surrounding hot gas pressure (blue), and cloud thermal to turbulent pressure ratio (red). Clouds are under-pressured when compared to their hot environment, and are typically supported more by their turbulent pressure.}
    \label{fig:clouds_pressratio_hist}
\end{figure}

Fig.~\ref{fig:slices_plot} shows that the morphology of clouds is generally not spherical, but rather elongated in the direction of the outflow. To quantify this elongation, we define the poloidal spread $\sigma_{\mathrm p}$ as the root-mean-square extent of the cloud in the poloidal plane (radial and vertical directions), computed as the trace of the covariance matrix of the cell positions relative to the cloud centroid. The toroidal spread $\sigma_{\mathrm t}$ is defined similarly, using deviations along the toroidal (azimuthal) direction. 
These measures provide a simple way to capture the cloud’s shape along and across the outflow. The poloidal elongation is then defined as the ratio $\sigma_{\rm p}/\sigma_{\rm t}$, so that a perfectly spherical cloud has a value of 1, while a cloud elongated in the $z$ or poloidal direction has a value greater than 1. Using this definition, in Fig. \ref{fig:clouds_dens_vel} (third panel) we show the poloidal elongation of the clouds as a function of the absolute velocity difference between them and the surrounding hot wind. Most clouds are indeed preferentially elongated in the poloidal direction, aligned with the wind direction. 
We also observe a clear trend of increasing elongation with velocity difference: clouds experiencing a stronger relative wind (in their rest frame) become more elongated, as they are subject to more effective ram pressure and hydrodynamical instabilities.

Finally, we compare the average thermal pressure $P_\mathrm{th} = n k_{\mathrm{B}}T$ within each cloud to that of the surrounding hot wind. The blue histogram of Fig.~\ref{fig:clouds_pressratio_hist} shows the distribution of cloud to wind pressure ratio, $P_\mathrm{cl, th} / P_\mathrm{w, th}$: \hi\ clouds are consistently under-pressured relative to the hot gas (i.e., $P_\mathrm{cl, th} / P_\mathrm{w, th} < 1$), which can also be appreciated in the slice map of Fig.~\ref{fig:slices_plot} (third panel). 
We also quantify the turbulent pressure in each cloud as $P_\mathrm{cl,turb}$ = $\rho \sigma_\mathrm{v}^2$, where $\sigma_\mathrm v = (\sigma_\mathrm{v,x}^2 + \sigma_\mathrm{v,y}^2 + \sigma_\mathrm{v,z}^2)^{1/2}$ is the 3D cloud velocity dispersion. We find that clouds are generally dominated by turbulent pressure. This is illustrated clearly in the red histogram of Fig. \ref{fig:clouds_pressratio_hist}, which shows the ratio $P_\mathrm{cl, th} / P_\mathrm{cl, turb}$: for $\sim80$\% of the cloud population, this ratio is below unity. We also estimate the magnetic pressure $P_\mathrm{cl, B} = B^2/8\pi$ within the clouds and find it to be typically at least two orders of magnitude lower than the thermal pressure, and often even smaller ($P_\mathrm{cl, th} / P_\mathrm{cl, B} \sim 10^2-10^6$). These results indicate that turbulence provides significant support against the external pressure exerted by the hot wind, whereas magnetic fields do not.

\subsection{Cloud properties as a function of distance}
\label{sec:clouds_zprofs}

\begin{figure*}[tp]
    \centering
    \includegraphics[width=0.95\textwidth, trim = 0 0 0 2cm, clip]{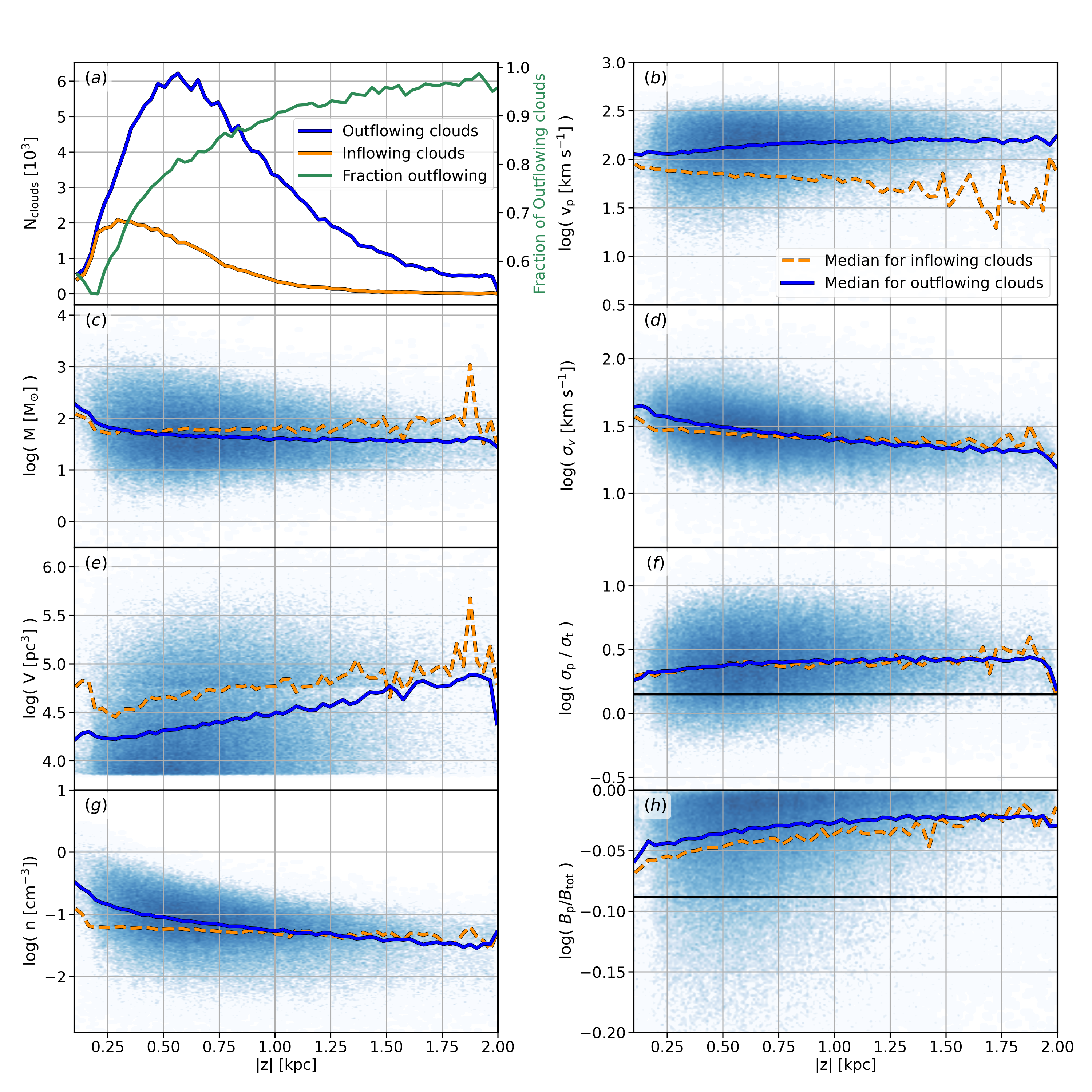}
    \caption{Properties of extraplanar \hi\ clouds as a function of their distance from the mid-plane $|z|$. Left column, from top to bottom: 
    number of outflowing and inflowing clouds together with the fraction of outflowing clouds $(a)$, mass $(c)$, volume $(e)$ and number density $(g)$.
    Right column: poloidal velocity $(b)$, 3D velocity dispersion $(d)$, poloidal elongation  (ratio of poloidal to toroidal spreads $\sigma_\mathrm{p} / \sigma_\mathrm{t}$) $(f)$ and magnetic poloidal alignment (ratio of poloidal to total magnetic field components $B_\mathrm{p}/B_\mathrm{tot}$) $(h)$. Blue colour maps denote the distributions of outflowing clouds, with blue solid lines representing the corresponding medians; dashed orange lines denote the medians for inflowing clouds. Black horizontal lines represent the spherical limit $\sigma_\mathrm{p} = \sqrt{2} \sigma_\mathrm{t}$ in panel $(e)$, and random alignment of the magnetic field ($B_\mathrm{p}^2 = \frac{2}{3}B_\mathrm{tot}^2$) in $(h)$.
    }
    \label{fig:clouds_zprofs}
\end{figure*}

We now look at the evolution of cloud properties as a function of $|z|$, i.e.\ their vertical distance from the Galactic plane, differentiating between inflowing and outflowing clouds.
Panel $(a)$ of Fig. \ref{fig:clouds_zprofs} shows the number of clouds that are outflowing (blue) and inflowing (orange) as a function of $|z|$, as well as the outflowing fraction (green). While at low $|z|$ there is more or less equal numbers of inflowing and outflowing clouds, at $|z|\gtrsim500$~pc most clouds are outflowing, and globally the outflowing fraction increases with height, reaching nearly unity at 2~kpc. 
At least a fraction of the extra-planar cool gas follows a galactic fountain trajectory, in which clouds are continuously expelled from the disc but rapidly fall back without reaching large heights (92.3\% of inflowing clouds are found at $|z|\lesssim1$ kpc). 
However, because the vast majority of cool gas that reaches $|z|\gtrsim1$ kpc is observed to be outflowing, a significant fraction of clouds are also unlikely to be participating in this galactic fountain flow.

The remaining panels in Fig. \ref{fig:clouds_zprofs} show the variation of several key cloud properties as a function of $|z|$: poloidal velocity (panel $b$), mass $(c)$, 3D velocity dispersion $(d)$, volume $(e)$, poloidal elongation $(f)$, density $(g)$ and poloidal magnetic field alignment $(h)$. These quantities are derived within each cloud before being averaged over the whole cloud sample, so that the density profile shown in panel $(g)$ represents the typical density of clouds and is not directly comparable to Fig. \ref{fig:zprofs_piernik}$a$.
Each panel shows the joint distribution of the outflowing clouds in the $|z|$–property plane, with colour indicating the number of gas cells per bin.
The blue thick curves denote the conditional medians for outflowing clouds, while the orange dashed lines indicate the corresponding medians for inflowing clouds.
Fig. \ref{fig:clouds_zprofs} highlights how, although at a given height the distributions of cloud properties are broad and can span several orders of magnitude, clear evolutionary trends emerge in their median values.
As before, we focus here on outflowing clouds, while inflowing objects will be discussed in Section~\ref{sec:clouds_infalling}.

From panel $(b)$, the median cloud velocity increases moderately with $|z|$, rising from $\sim 100~\kms$ in the proximity of the disc to about $150~\kms$ at $|z|\sim2$ kpc.
This is suggestive of continuous acceleration of cool clouds by the ambient hot flow as they propagate through the wind.
The cloud Mach number (not shown here) also increases with $|z|$, driven both by the rise in cloud velocity and by a decline in the hot gas temperature at higher heights, resulting from a combination of adiabatic expansion and mixing with the warmer gas. Consequently, clouds transition from predominantly subsonic motion ($\mathcal{M} < 1$) at low $|z|$ to nearly transonic or mildly supersonic motion ($\mathcal{M}\sim 1$) at large $|z|$.
In addition, the cloud velocity dispersion $(d)$ clearly declines with $|z|$, decreasing by roughly a factor of two over 2 kpc in height (from $\sim 30 ~\kms$ to $15~\kms$).

The cloud masses $(c)$ exhibit a mild but systematic decline, i.e. clouds tend to be more massive closer to the disc than away from it.
The median mass of outflowing clouds decreases by more than a factor of two over 2 kpc, from $\sim 2000 ~\Msun$ close to the disc to $\sim 800 ~\Msun$ at $|z|\sim2$ kpc. This trend may be attributed to mass loss through mixing with the hot gas, combined with the reduced acceleration of more massive clouds to large heights.
Mixing with the hot phase is indeed visible in Fig. \ref{fig:display_phases} (third panel), as the intermediate temperatures between the cool clouds and the hot wind (between $\sim2\times10^4$ K and $10^6$~K) are found out of the disc only around cloud locations. Furthermore, the temperature map of Fig. \ref{fig:slices_plot} (second panel) shows that clouds are embedded in gas at higher temperature ($\sim10^5$~K) and leave a trail of gas with temperatures between 10$^5$~K and 10$^6$~K.

While their mass decreases, clouds tend to grow in volume with $|z|$ $(e)$.
This trend is likely linked to interactions and mixing with the surrounding hot medium: as clouds rise to larger heights, they become increasingly dispersed and elongated due to shear and stripping by the hot wind. This is consistent with panel $(f)$, showing the poloidal elongation growing with height.
In addition, because the thermal pressure of hot gas decreases with $|z|$, clouds at larger heights are expected to expand in response to the lower ambient pressure. Note that the sharp cut-off below volumes $\sim$10$^{3.8}$~pc$^3$ is due to the removal of all clouds with less than 4 cells per cloud radius, as described in Section~\ref{sec:Clouds}.
The simultaneous decrease in mass and increase in volume implies a corresponding decline in cloud density $(g)$.
The median density drops by more than an order of magnitude, from $\sim 0.3$ cm$^{-3}$ close to the disc to $\sim 0.02$ cm$^{-3}$ at $|z|\sim2$ kpc.
The density contrast between the cool and hot surrounding gas ($\chi = n_\mathrm{cool} / n_\mathrm{hot}$) is fairly constant with height and typically of the order of $10^2$.
Panel $(h)$ is motivated by the apparent alignment of the magnetic field with the wind direction, as noted in Section~\ref{sec:Clouds} and Fig.~\ref{fig:slices_plot}. To quantify this more systematically, we compute for each cloud the ratio $B_\mathrm{p}/B_\mathrm{tot}$, i.e. the poloidal component of the magnetic field relative to its total strength, as a function of $|z|$. Above the disc, the magnetic field is preferentially oriented in the poloidal direction, i.e. parallel to the cloud shear surface, with this alignment strengthening at larger heights. A similar trend is also found in the hot gas phase, indicating that the magnetic field lines become progressively aligned with the outflow velocity outside the disc. The implications of this alignment for cloud survival are discussed in Section~\ref{sec:clouds_survival}.

Overall, the thermodynamical and kinematical quantities are consistent with a scenario in which cool gas clouds are lifted from the Galactic plane and entrained in the hot wind.
As shown in previous works \citep[e.g.,][]{Schneider+20, Tan+24, Ramesh26}, such clouds can gain momentum and can be accelerated during their ascent through a combination of drag force exerted by the hot phase and mixing with it.
As they propagate, clouds evolve with distance by mixing with the ambient medium: they become elongated and more diffuse, while a fraction of the atomic gas turns into ionized material, leading to a net loss of atomic mass. 
In this context, the most challenging trend to interpret is the decline of velocity dispersion, since one might naively expect enhanced turbulence as cool clouds mix with the hotter phase.
One possible, speculative interpretation is that, while the colder inner cores of the clouds become more homogeneous over time, the warmer and more turbulent outer envelopes are progressively ionized and dispersed into the hot wind, and therefore no longer contribute to the measured velocity dispersion.

\subsection{Cloud properties as a function of time and SN activity}
\label{sec:clouds_time}

\begin{figure*}[tp]
   \centering
    \includegraphics[width=0.95\textwidth]{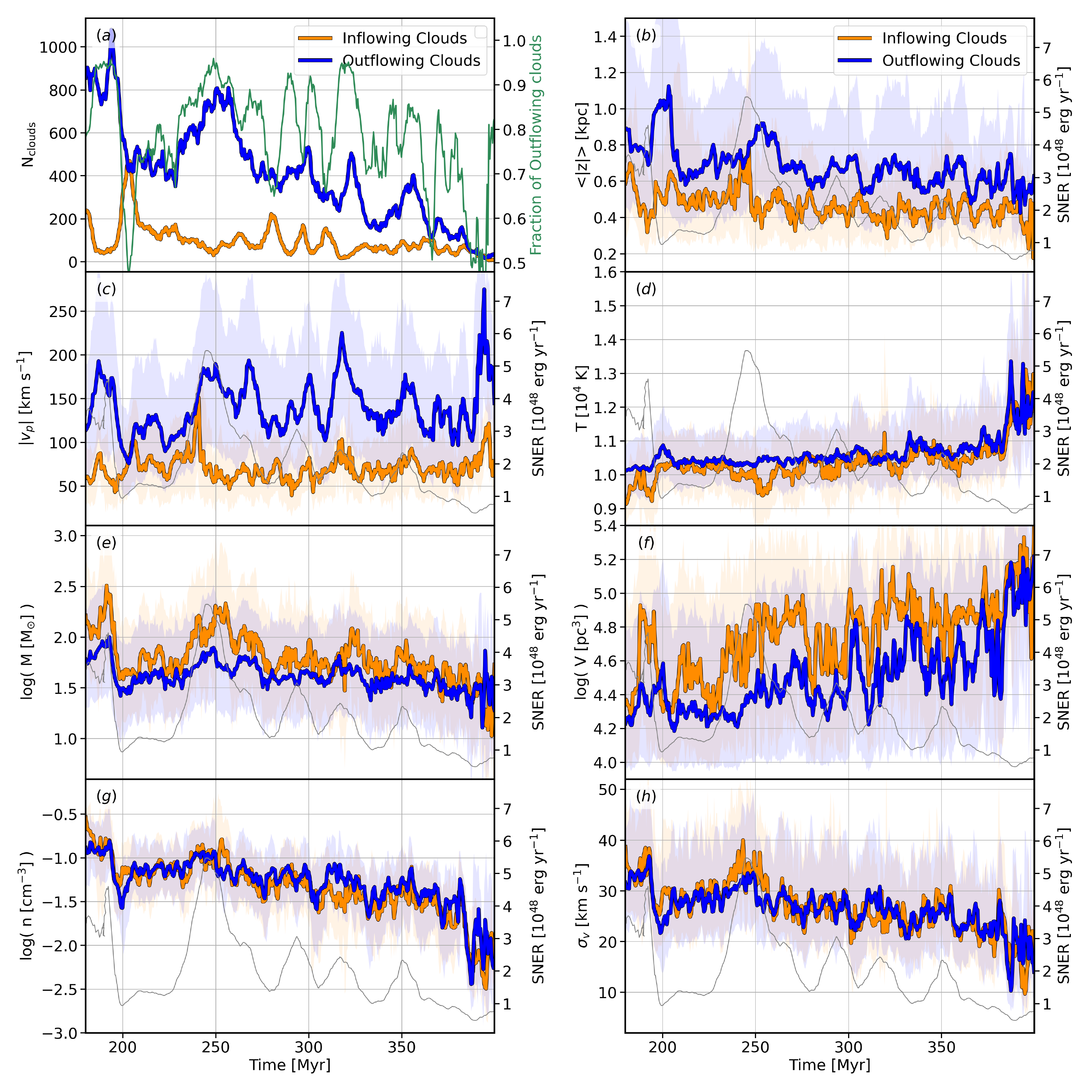}
    \caption{Properties of outflowing (blue) and inflowing (orange) clouds as a function of time, together with the supernova energy rate (thin grey curve, moving-average over 10~Myr). For each property, we measure the median of the distributions (solid lines), and the 16$^{\rm th}$ and 84$^{\rm th}$ percentile (shaded areas). Left column, from top to bottom: number of clouds together with outflowing fraction in green $(a)$, poloidal velocity $(c)$, mass $(e)$ and number density $(g)$. Right column: median distance from the mid-plane $\langle \,|z| \, \rangle$ $(b)$, temperature $(d)$, volume $(f)$ and velocity dispersion $(h)$.}
    \label{fig:clouds_time}
\end{figure*}

\begin{figure*}[tp]
   \centering
    \includegraphics[width=0.95\textwidth]{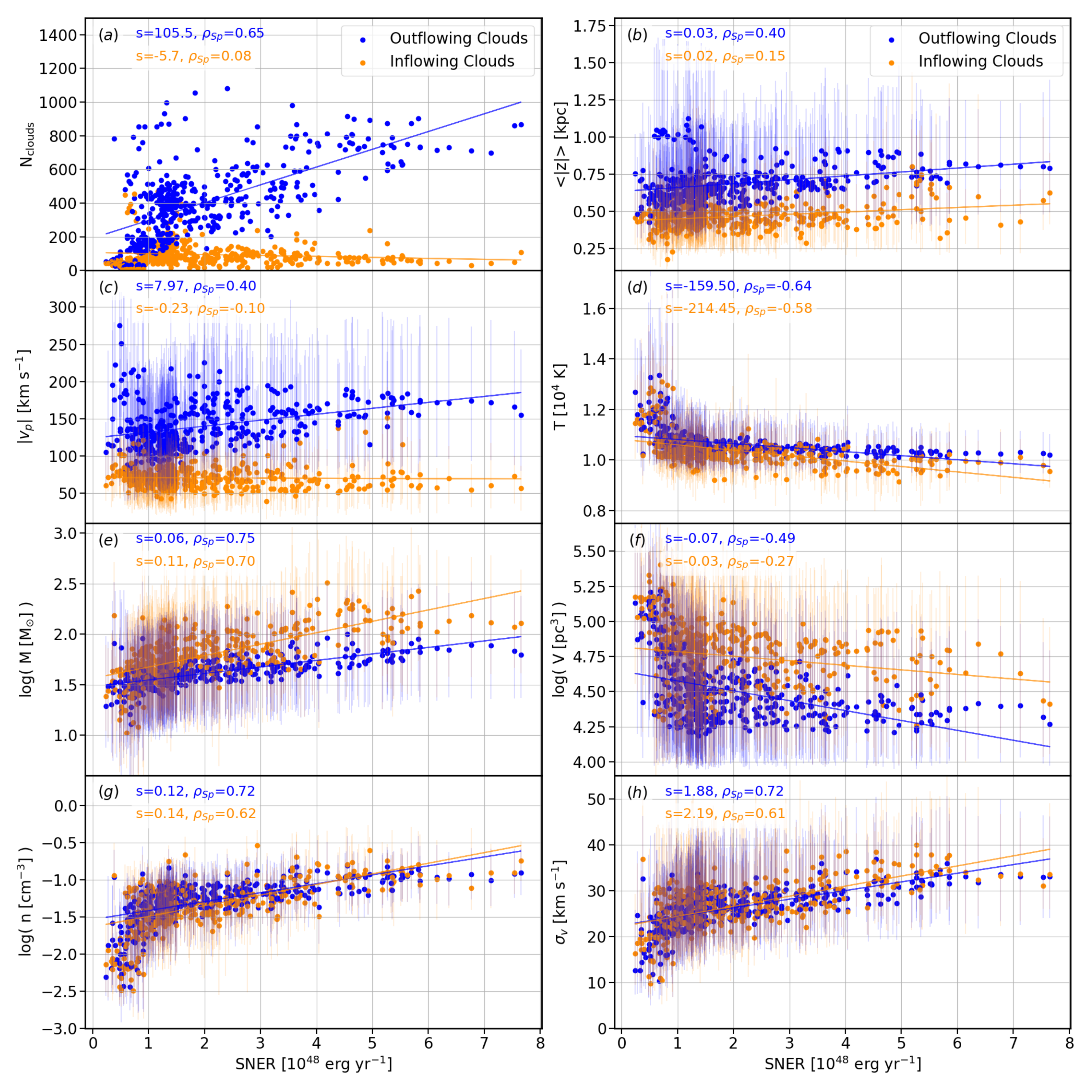}
    \caption{Properties of outflowing (blue) and inflowing (orange) clouds as a function of SNER. Points represent the median values in different snapshots of the simulation, with the 16$^{\rm th}$ and 84$^{\rm th}$ percentiles indicated as error bars. Left column, from top to bottom: number of clouds $(a)$, poloidal velocity $(c)$, mass $(e)$ and density $(g)$. Right column: median distance from the mid-plane, $(b)$, temperature $(d)$, volume $(f)$ and velocity dispersion $(h)$. The solid lines represent linear regressions separately for outflowing and inflowing clouds, with their best-fit slopes $s$ indicated at the top of each panel, together with the Spearman correlation coefficient $\rho_\mathrm{Sp}$.} 
    \label{fig:clouds_sner}
\end{figure*}

Outflows in our simulation are driven by stellar activity, which varies significantly over time, as discussed in Section~\ref{sec:gas_dynamics}. As a consequence, cloud properties are expected to also vary over the course of the simulation, reflecting changes in the underlying star formation and SN activities.
In this section, we examine how the properties of simulated clouds change as a function of time and SNER.
Fig. \ref{fig:clouds_time} summarizes the evolution of the median cloud properties throughout the simulation time: we analyse the number of clouds (panel $a$), their median absolute height $(b)$, poloidal velocity $(c)$, temperature $(d)$, mass $(e)$, volume $(f)$, number density $(g)$ and velocity dispersion $(h)$.
As before, the outflowing sample is shown in blue, while the inflowing sample is in orange. To visualize the potential link to SN activity, we also plot the SNER as a thin gray line. 
In Fig.~\ref{fig:clouds_sner}, we explicitly plot the same quantities as a function of SNER and investigate correlations with it. We note that the SNER and cloud properties are measured at the same instant, without accounting for the travel time of clouds. Objects at large $|z|$ are more likely tied to past SN activity from when they originally left the Galactic mid-plane.

The number of outflowing clouds (Fig. \ref{fig:clouds_time}$a$) clearly decreases over time, dropping from more than 2000 at the beginning of the simulation to only $\sim$100 by its end. This long-term decline mirrors the systematic decrease in SN activity throughout the simulation, as discussed in Section~\ref{sec:gas_dynamics}.
Beside this global trend, short-timescale fluctuations closely track variations in the SNER (e.g., peaks around 250 Myr and 350 Myr).
As expected, enhanced SN activity, following enhanced star formation activity, accelerate a larger number of clouds out of the disc.
Such a correlation clearly appears also in Fig. \ref{fig:clouds_sner}$a$, where the number of outflowing clouds shows an increasing trend as a function of SNER, with a Spearman correlation coefficient of $\rho_\mathrm{Sp}=0.65$.
In contrast, the number of inflowing clouds shows a much milder decrease over time and a weaker correlation with the SNER ($\rho_\mathrm{Sp}=0.08$).
Nevertheless, we can note that short-timescale peaks in the number of inflowing clouds tend to occur during periods of low SN activity (e.g., around 210 Myr and 280 Myr), while local minima often coincide with peaks in star formation and feedback activity (e.g., around 250 Myr and 320 Myr).
This behaviour suggests that clouds are more likely to fall back onto the disc when stellar feedback is less effective. 
As a result of these trends, the fraction of outflowing clouds, shown in green in Fig. \ref{fig:clouds_time}$a$, fluctuates over time, largely tracking oscillations in the number of both outflowing and inflowing clouds. This fraction links to the SNER, with periods of more intense star formation leading to higher fractions of outflowing clouds.

Panels $(b)$ of Figs.~\ref{fig:clouds_time}-\ref{fig:clouds_sner} show the median absolute distance from the mid-plane $\langle\,|z|\,\rangle$ of the cloud population. For the outflowing sample, this quantity slightly decreases over time, from $\sim 0.8$ kpc at early times to $\sim 0.5$ kpc at late times. 
As for the number of clouds, this decrease of $\langle\,|z|\,\rangle$ is related to the declining star formation activity over the course of the simulation and it seems to broadly follow the evolution of the SNER, showing a moderate correlation with it ($\rho_\mathrm{Sp}=0.40$), meaning that stronger feedback can drive clouds further away from the Galactic plane. 

The cloud masses and densities are globally declining over time (Fig. \ref{fig:clouds_time}$e$-$g$), because the decreasing SN activity becomes less effective at accelerating dense, massive clouds out of the disc. This is supported by the strong positive correlations of mass and density with the SNER seen in panels $(e)$ and $(g)$ of Fig. \ref{fig:clouds_sner} ($\rho_\mathrm{Sp}=0.75$ and 0.72 respectively). Conversely, cloud temperatures increase over time (Fig. \ref{fig:clouds_time}$d$), which corresponds to a decreasing trend with SNER (Fig. \ref{fig:clouds_sner}$d$, $\rho_\mathrm{Sp}=-0.64$).
This is consistent with a scenario in which clouds are entrained and lifted out of the disc by the hot wind, with higher SN activity producing winds capable of accelerating more massive clouds, composed of denser and colder gas. Towards later times, the SNER is weaker, which leads to lighter and marginally warmer outflowing clouds.
While the cloud volumes show a slight increase over time (Fig. \ref{fig:clouds_time}$f$), we only find a moderate anti-correlation with SNER (Fig. \ref{fig:clouds_sner}$f$, $\rho_\mathrm{Sp}=-0.49$). 
High SN activity lifts denser and more massive clouds, but not necessarily larger ones. 
The temporal increase in cloud volume is likely driven by the decline in the hot-phase temperature of the outflows: as SN activity decreases, the thermal pressure of the hot wind drops, allowing clouds to expand. In fact, the cloud density decreases more rapidly with time than the cloud mass, supporting this interpretation.

The cloud poloidal velocities vary over time (Fig. \ref{fig:clouds_time}$c$) and broadly follow the SN activity. Looking at panel $(c)$ of Fig. \ref{fig:clouds_sner}, we find that the velocities indeed tend do increase with the SNER, with a Spearman correlation coefficient of $\rho_\mathrm{Sp}=0.40$. This is consistent with the moderate correlation found for $\langle\,|z|\,\rangle$ as a function of the SNER (Fig. \ref{fig:clouds_sner}$b$): during periods of intense SN activity, clouds are accelerated to higher velocities, therefore reaching greater distances from the mid-plane. Interestingly, the correlation for these two properties with SNER is weaker than for other quantities, which may be related to the tendency of higher-density clouds to move more slowly, as shown previously in Fig. \ref{fig:clouds_dens_vel} (left panel). In periods of high SNER, clouds tend to be accelerated more, but this effect is moderated by the population of clouds being denser and more massive, as established above (Fig. \ref{fig:clouds_sner}$e$-$g$), and therefore slower. In order to remove this bias, we checked the correlation between velocity and SNER dividing clouds in density bins, and generally found higher Spearman correlation coefficients (e.g., $\rho_\mathrm{Sp}=0.52$ for cloud densities of 0.2~cm$^{-3}$). In addition, it is worth noting that the weak correlation between outflow velocity and SFR has been found previously both in simulations and in observations, and is generally explained in terms of weak dependence of the velocity of the SN-driven shell on the SFR \citep[e.g.][]{2010AJ....140..445C,2020ApJ...900...61K}.
Finally, the velocity dispersions are globally declining over time (Fig. \ref{fig:clouds_time}$h$), with peaks in periods of intense SN activity. Consistently, they show a clear positive trend with SNER ($\rho_\mathrm{Sp}=0.72$), as seen in Fig. \ref{fig:clouds_sner}$h$.
This evidence can be interpreted as the result of enhanced turbulence in the ISM driven by frequent SN events \citep[e.g.,][]{2016MNRAS.458.1671K}, which in turn may increase the internal velocity dispersion of the outflowing clouds. Furthermore, the subsequent winds are stronger and more disruptive for cold clouds, also augmenting their internal turbulence.

In summary, the properties of outflowing clouds are tightly coupled to the time-variable star formation and SN activity in the CMZ. Enhanced feedback increases the number of clouds launched into the outflow. During these phases, not only low-density clouds but also more massive and denser structures are accelerated to higher velocities. As the SN activity declines, the outflows weaken, fewer clouds are lifted from the disc, and the cloud population becomes progressively lighter, warmer, and less turbulent. 
Taken together, the findings presented so far strongly support entrainment by the SN-driven hot wind as the dominant cloud acceleration mechanism \citep[e.g.][]{Schneider+18,Fielding+2022,Tan+24}, rather than, e.g., clouds or filaments spontaneously forming via thermal instability in the wind \citep{2012MNRAS.419.3319M,2016MNRAS.455.1830T, Nguyen2024}.

\subsection{Fountain flows: infalling clouds}
\label{sec:clouds_infalling}

In this section, we examine the properties of inflowing clouds and compare them to the outflowing sample. At all times in the simulation, the number of outflowing clouds largely exceeds that of inflowing ones (Fig.~\ref{fig:clouds_time}$a$), indicating that a significant fraction of objects does not return to the disc in a fountain cycle. The fraction of outflowing clouds, while fluctuating with the SNER, hovers around 80\%. This implies a lower limit of 20\% for clouds that are recycled into fountain flows, since inflowing clouds are likely to have already been counted as outflowing at earlier snapshots. As expected, the median height $\langle\,|z|\,\rangle$ is larger for the outflowing population than for the inflowing one at all times (Fig.~\ref{fig:clouds_time}$b$), consistent with the fraction of outflowing clouds increasing with $|z|$ (Fig. \ref{fig:clouds_zprofs}$a$). Together, these trends suggest that the majority of outflowing clouds, up to $\sim$80\%, either leave the disc permanently or are destroyed during their ascent.

Infalling clouds are, on average, characterized by lower absolute velocities than outflowing ones, with a median inflow velocity of 74~$\kms$, compared to 139~$\kms$ for the outflowing population (Fig.~\ref{fig:clouds_time}$c$). The different median velocities reflect the fact that only relatively low-velocity clouds return to the disc, while higher-velocity clouds continue outward and do not contribute to the infalling population. The velocities of the infalling clouds remain relatively stable over time, with much smaller fluctuations than their outflowing counterparts.
The cloud velocities decrease with $|z|$ (Fig. \ref{fig:clouds_zprofs}$b$), consistently with a ballistic trajectory in a fountain flow, in which clouds reach zero velocity near the apex and then are gradually accelerated as they fall back toward the disc. Nevertheless, these clouds typically have velocities lower than the free-fall velocity expected for clouds falling from any height above 500~pc, due to the deceleration effect of the outflowing hot wind.

The remaining properties of infalling clouds, i.e. temperature, mass, density, volume and velocity dispersion (Fig. \ref{fig:clouds_time}$d$-$f$-$g$-$h$), appear very similar to their outflowing counterparts and follow a similar time evolution. However, this similarity is partially driven by a bias arising from the systematically lower median height $\langle\,|z|\,\rangle$ of inflowing clouds compared to outflowing ones. Since cloud properties vary with $|z|$ (Section \ref{sec:clouds_zprofs}), a direct comparison between populations occupying different $|z|$ ranges can be misleading. To mitigate this effect, we restrict the comparison to clouds with $|z|<600$~pc. 
This cut includes most inflowing clouds but less than half of the outflowing sample (Fig. \ref{fig:clouds_zprofs}$a$), leaving the median inflowing properties largely unchanged while affecting some properties of the outflowing clouds.
For the latter, the height cut has a negligible impact on temperature, but increases the median mass, density, and velocity dispersion (by factors of $\sim$1.2, $\sim$1.7 and $\sim$1.2, respectively) and decreases their velocity and volume (by factors of $\sim$1.1 and$\sim$1.5), in agreement with the trends in Fig. \ref{fig:clouds_zprofs}. 
After applying the $|z|<600$~pc cut, inflowing clouds have, on average, lower masses, densities and velocity dispersions, and larger volumes than outflowing ones. At a first glance, this may appear counter-intuitive, as one might expect the most dense and massive clouds to be those falling back onto the disc, due to the stronger gravitational pull. However, this behaviour can be explained if clouds gradually lose mass to the hot wind over time, as shown in Section~\ref{sec:clouds_zprofs}. In this picture, infalling clouds have experienced more prolonged interaction with the hot phase, leading to greater elongation and mass loss, and resulting in larger volume, lower density and declining velocity dispersion. This interpretation is further supported by panel $(c)$ of Fig. \ref{fig:clouds_zprofs}, which shows a mild increase in inflowing cloud mass with $|z|$: clouds at lower heights are closer to the end of their ballistic trajectories, implying longer exposure to the wind and hence more advanced mass loss.

Most properties of inflowing clouds follow trends with SNER similar to those of their outflowing counterparts (Fig. \ref{fig:clouds_sner}), albeit typically with slightly lower Spearman correlation coefficients. Likewise, in Fig. \ref{fig:clouds_zprofs}, the dependences of cloud volume, poloidal elongation, temperature and velocity dispersion on $|z|$ are comparable for both samples. This overall similarity confirms that inflowing and outflowing clouds belong to a single population, with differences in their properties arising primarily from the longer interaction time of inflowing clouds with the hot wind along their fountain trajectory. The main difference in Fig. \ref{fig:clouds_sner} is that neither the number (panel $a$) nor the velocity $(c)$ of infalling clouds shows a significant correlation with SNER ($\rho_\mathrm{Sp}=$0.08 and -0.10 respectively), indicating that their motion is governed by the gravitational pull rather than by the hot wind.
Although one might expect more inflowing clouds in periods of elevated SN activity, given the larger number of outflowing ones, the fraction of outflowing clouds is not constant over time or with SNER, as discussed in Section~\ref{sec:clouds_time}. Intense star formation produces a larger fraction of clouds leaving the disc without promptly returning, or returning only on longer timescales, possibly in more quiescent periods. This view is also supported by the increasing trend of $\langle \, |z|\, \rangle$ with SNER: stronger SN activity pushes clouds to larger distances, making it less likely for them to fall back on short timescales.

In conclusion, while many clouds are accelerated out of the disc, only a minority return on fountain trajectories, with the remainder either reaching larger heights or being disrupted during the acceleration. The systematic differences between the two populations at $|z|<600$ pc, mainly in mass, volume, density, and velocity dispersion, are naturally explained by the longer interaction time of inflowing clouds with the hot wind, leading to progressive mass loss and structural evolution.

\subsection{Cloud survival}
\label{sec:clouds_survival}

\begin{figure}
     \includegraphics[width=0.5\textwidth,trim = 0.5cm 0 0 0.1cm, clip]{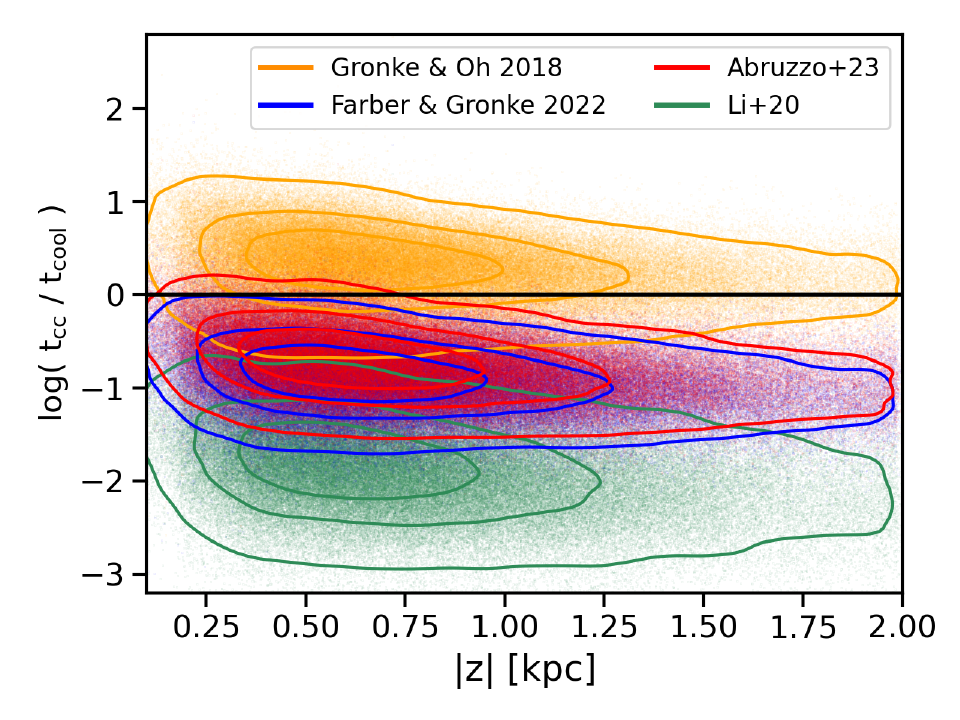}
     \caption{Ratio of cloud crushing time $t_{\mathrm{cc}}$ to cooling time $t_{\mathrm{cool}}$ for outflowing clouds as a function of distance from the mid-plane $|z|$. Different definitions of $t_{\mathrm{cc}}$ and the cooling time $t_{\mathrm{cool}}$ are explored here, corresponding to different works (see text), and are shown with different colours and contour lines. The black dotted horizontal line represents $t_{\mathrm{cc}}=t_{\mathrm{cool}}$: clouds situated above this line are considered as surviving.}
     \label{fig:clouds_survival}
\end{figure}

Interactions with the surrounding, fast-moving hot medium tend to shred and evaporate cold gas on short timescales \citep[e.g.,][]{Scannapieco&Bruggen15,Schneider+17, Tan+24}. Several mechanisms have been proposed to enhance cloud longevity, including turbulent mixing combined with radiative cooling of the hot phase \citep{Gronke+20,Banda-Barragan+21,Kanjilal+2021}, magnetic fields \citep{Sparre+20,Gronnow+22}, or thermal conduction \citep{Bruggen&Scannapieco16,Armillotta+17}. 
Our simulation does not currently allow us to track individual clouds over time, so that it is not straightforward to infer whether a given cloud survives on a long timescale. However we can attempt to predict which clouds are more likely to survive, based on some of the works mentioned above.

Several studies have analysed the time evolution and survival of idealised cool gas clouds moving through a hot wind tunnel, highlighting in particular the importance of radiative cooling in cloud survival. 
These works established that a cloud can survive if the gas around it can cool faster than the cloud is destroyed, allowing the cloud to gain mass more rapidly than it loses it to the wind. For example, this process was described in \cite{Gronke+19,Gronke+20}, who assumed that the relevant cooling time is set by the mixed phase forming at the cloud–wind interface, characterized by a temperature $T_{\mathrm{mix}}=\sqrt{T_{\mathrm{cool}} T_{\mathrm{hot}}}$, where $T_{\mathrm{cool}}$ and $T_{\mathrm{hot}}$ are the cloud and wind temperatures, respectively. The dynamical timescale of cloud destruction by the wind, also called cloud-crushing time, is defined as $t_{\mathrm{cc}}=\sqrt{\chi}r_{\mathrm{cl}}/v_{\mathrm{w}}$, with $r_{\mathrm{cl}}$ being the cloud radius, $\chi$ the density contrast, and $v_{\mathrm{w}}$ the hot wind velocity. Therefore, according to this study, a cloud survives if

\begin{equation}
t_{\mathrm{cool}} \equiv  \frac{k_{\mathrm{B}} T_{\mathrm{mix}}}{n\Lambda(n,T_{\mathrm{mix}})} < t_{\mathrm{cc}} ~~ .
\end{equation}

\noindent Later, more studies attempted to refine this criterion, often by changing the definitions of $T_{\mathrm{mix}}$ or of $t_{\mathrm{cc}}$. In particular, \citealt{Li+20} found better results when using directly $T_{\mathrm{hot}}$ instead of $T_{\mathrm{mix}}$ (which considerably increases $t_{\mathrm{cool}}$), and used, instead of $t_{\mathrm{cc}}$, a predicted cloud lifetime $t_{\mathrm{life,pred}}=10 \,t_{\mathrm{cc}} \tilde{f} $, with $\tilde{f} $ being a function of the cloud length, density and velocity. \cite{Farber&Gronke+22} then suggested to keep $t_{\mathrm{cc}}$ as defined by \cite{Gronke+19} but to redefine the temperature of the mixing layer as  $T_{\mathrm{mix}}=\sqrt{T_{\mathrm{min,cool}}T_{\mathrm{hot}}} $, where $T_{\mathrm{min,cool}}$ is the temperature comprised between $T_{\mathrm{cool}}$ and $T_{\mathrm{hot}}$ that minimises $t_{\mathrm{cool}}$. Shortly after, \cite{2023arXiv230703228A}, keeping the definition of $t_{\mathrm{cool}}$ introduced by \cite{Farber&Gronke+22}, argued that the relevant comparison timescale is not the cloud-crushing time $t_\mathrm{cc}$, but rather $\alpha t_{\mathrm{shear}}$,  the time required for wind fluid elements to cross the full cloud length. In this framework, clouds are expected to survive if
$t_{\mathrm{cool}} < \alpha t_{\mathrm{shear}}$, with $t_{\mathrm{shear}} = r_{\mathrm{cl}}/v_{\mathrm{w}}$, and $\alpha=7$ a constant that they obtain from a best fit to their data.

By applying these different criteria to the clouds found in our simulation, and using, like in Section~\ref{sec:clouds_general}, the properties of the hot wind ($T>10^6$~K) around each cloud within a box three times the size of the cloud, we find that the inferred survival fraction is highly sensitive to the chosen criterion. This is expected at large $\chi$ values ($\sim100$ for our clouds), where the criteria for cloud survival are uncertain. Applied to all of our 175077 outflowing clouds, we find that the \citet{Gronke+20}'s criterion predicts that 78.4\% of them survive, but this number drops to 0.5\% for the \citet{Farber&Gronke+22}'s criterion, to 2.0\% for \citet{2023arXiv230703228A}'s, and to 0.03\% for \citet{Li+20}'s.
In Fig. \ref{fig:clouds_survival}, we plot the four $t_{\mathrm{cc}}/t_{\mathrm{cool}}$ criteria as a function of height for outflowing clouds. When going towards higher $|z|$, the $t_{\mathrm{cc}}/t_{\mathrm{cool}}$ ratio decreases, implying a decrease in the fraction of surviving clouds, regardless of the criterion chosen. This is expected, as cloud densities decrease with $|z|$ while their volume increases due to elongation and shredding (Section \ref{sec:clouds_zprofs}). 
We also investigated whether survival fractions depend on time or the SNER. However, we found no significant trends, with the survival fractions remaining roughly constant throughout the simulation. 

For inflowing clouds, we find the \citet{Gronke+20}, \citet{Farber&Gronke+22}, \citet{2023arXiv230703228A} and \citet{Li+20} criteria to give survival predictions of 89.5\%, 2.7\%, 7.2\% and 0.05\%, respectively. The survival rates are therefore higher for inflowing than for outflowing clouds. This, together with the fact that we have consistently more outflowing clouds than inflowing ones throughout the simulation (by a factor of $\sim$4, Section~\ref{sec:clouds_infalling}), confirms that some outflowing clouds survive and are recycled into fountain flows, while others dissolve into the hot wind before being able to fall back onto the disc.
Although clouds that rise above $|z|=2$~kpc are no longer tracked by the AMR, only a small minority of clouds actually reach such heights, as shown in Fig. \ref{fig:clouds_zprofs}$a$.
The fraction of clouds surviving while being accelerated to $|z|\gtrsim2$~kpc therefore appears negligible over the entire simulation, but can increase slightly during periods of strong star formation. 
Over the entire run, clouds at $|z|>1.9$~kpc account for only 1.8\% of the total cloud population, with this fraction varying between 0 and $\sim$5\% over time.

Although determining the most appropriate criterion for our clouds is not straightforward, we can use the fraction of inflowing clouds as an empirical proxy for cloud survival. Because very few objects reach heights larger than $\sim$2~kpc, the most likely fate of a cloud after leaving the disc is either destruction by the hot wind or fallback in a fountain flow. The fraction of inflowing clouds can therefore be regarded as an upper limit on the survival fraction. This allows us to rule out the \cite{Gronke+19} criterion for overestimating survival, as we clearly do not observe 78\% of outflowing clouds being recycled into fountain flows (see Section~\ref{sec:clouds_infalling}). By contrast, the remaining three criteria, which predict survival fractions below 2\%, are more consistent with our results but would imply that only a small fraction of clouds survive long enough to return to the mid-plane. 

Several studies have shown the importance of magnetic field for clouds evolution and survival \citep[e.g.][]{Gronke+20, Li+20, Sparre+20, Kaul+25}. 
In our simulation, as mentioned in Section \ref{sec:clouds_general}, the thermal pressure within \hi\ clouds is at least two orders of magnitude higher than the magnetic pressure, implying negligible contribution of magnetic pressure to the cloud support.
While the alignment of the cloud magnetic field with the wind direction, found for most clouds in this simulation, might suggest that a parallel magnetic field suppresses or weakens the development of Kelvin-Helmholtz instabilities at the cloud-wind interface, thereby increasing the cloud stability against destruction \citep[e.g.,][]{Bruggen+23}, we find that the stability criterion \citep{1961hhs..book.....C} is largely unmet: the Alfvén speed both around clouds and within them is much smaller than the relative cloud-wind velocity. 
Given the above, magnetic field is unlikely to play a significant role in the evolution and survival of our simulated clouds.

 \subsection{Comparison to \hi\ observations}
 \label{sec:clouds_obs}

 \begin{figure}
    \centering
    \includegraphics[width=0.49\textwidth, trim= 0.0cm 0 0 0.0cm, clip]{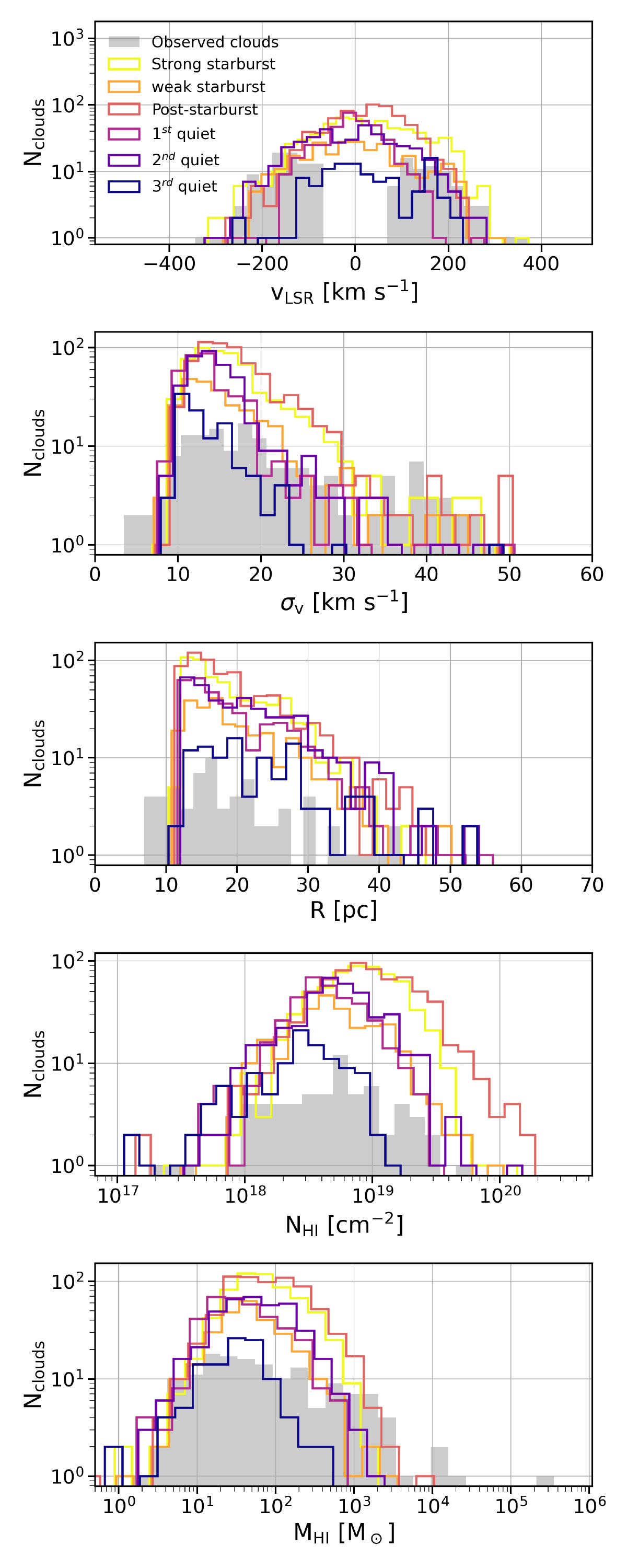}
    \caption{Properties of simulated clouds (coloured open histograms) compared to observed \hi\ clouds (grey histograms) from \cite{DiT+18}. From top to bottom: LSR velocities, velocity dispersions, radii, column densities and masses. We plot simulated populations at epochs with different star formation activities (see Figs. \ref{fig:sfr_sner} and \ref{fig:display_times}).}
    \label{fig:clouds_comp_obs}
\end{figure}

A few hundred high-velocity \hi\ clouds have been detected above and below the GC thanks to dedicated radio surveys \citep{McClure-Griffiths+13, DiT+18}.
These clouds have Local Standard of Rest (LSR) velocities reaching up to $\sim400~\kms$, and their kinematics are inconsistent with Galactic rotation. Instead, they are best explained by a bi-conical outflow launched from the GC and accelerating with distance \citep{2020ApJ...888...51L}.
Here we investigate whether our simulations produce a \hi\ cloud population similar to the observed one. 
Specifically, we compare our simulated clouds, at different epochs of the simulation, to the \hi\ data from \cite{DiT+18} taken with the Green Bank Telescope (GBT).
The observational sample consists of 155 clouds located approximately within a longitude $|\ell| \lesssim 10^\circ$ and a latitude $|b| \lesssim 10^\circ$, roughly corresponding to a 3.5 kpc $\times$ 3.5 kpc region around the GC.

In Fig. \ref{fig:clouds_comp_obs}, from top to bottom, we compare the distributions of LSR velocity $v_\mathrm{LSR}$, line-of-sight velocity dispersion $\sigma_\mathrm{LOS}$, radius $R_\mathrm{C}$, \hi\ column density $N_\hi$ and mass $M_\hi$. 
Filled grey histograms denote the observed cloud population, while coloured open histograms show the simulated population at the six times introduced in Section \ref{sec:Outflows}. 
Because observations cannot distinguish between inflowing and outflowing clouds, being the kinematic information limited to the line-of-sight velocity,
we include all simulated clouds in the comparison. 
Three-dimensional velocity vectors are converted into LSR velocities using the standard equation for an object seen at a given Galactic longitude $\ell$ and latitude $b$ \citep[e.g.,][]{DiT+18}:

\begin{equation}
    v_{\mathrm{LSR}} = \left(v_{\theta}\frac{R_{\odot}}{R} - v_{\odot}\right) \sin(\ell) \cos(b) + v_z\sin(b) - v_R\cos(\ell+\theta) \cos(b)
\end{equation}

\noindent where $R$ and $\theta$ are the cylindrical radius and the azimuthal angle of the object, respectively, ($v_R$, $v_\theta$, $v_z$) its velocity components in cylindrical coordinates. $R_{\odot}=8.2$~kpc and $v_{\odot}=240~\kms$ are the solar radius and tangential velocity, respectively \citep{Bland-Hawthorn+2016}.
The line-of-sight velocity dispersions are calculated within each cloud from the $v_{\mathrm{LSR}}$, adding also the thermal broadening, i.e.\ $\sigma_\mathrm{v}=\sqrt{\sigma_\mathrm{v, \, LSR}^2+c_s^2}$, with $c_s$ being the sound speed within the cloud.
The radii of simulated clouds are simply derived from their volume $V$, assuming a spherical shape, as $R_\mathrm{C} = (\frac{3}{4} V / \pi)^{1/3}$, while similarly the observed clouds radii are derived as $R_\mathrm{C} = \sqrt{(A/\pi)}$, $A$ being the area covered by the clouds onto the sky. The column densities of both observed and simulated clouds are derived as the masses divided by the assumed circular area, i.e.\ $N_\hi = M_\hi / (\pi R^2_\mathrm{C})$.

The LSR velocities of simulated clouds (Fig. \ref{fig:clouds_comp_obs}, top panel) are in good agreement with the observations for clouds with $|v_\mathrm{LSR}| \gtrsim 50~\kms$, reaching up to $\sim400~\kms$ and reproducing the observed decline in cloud numbers at the highest velocities. 
We note that the observed sample lacks low-velocity clouds ($|v_\mathrm{LSR}| \lesssim 50~\kms$). This absence is not physical, but instead reflects an observational bias: strong emission from local gas in the solar neighbourhood dominates the signal near $v_\mathrm{LSR}\simeq0~\kms$, causing potential low-velocity outflowing clouds to be confused with, and effectively obscured by, foreground emission. 
This produces the double-peaked velocity distribution that characterizes the observations. 
On the contrary, the distribution of simulated clouds peaks at $v_\mathrm{LSR}\simeq0~\kms$, implying that the bulk of the population is moving predominantly in the vertical direction and is located close to the axis of the outflow cone, where line-of-sight velocity projections are minimal.
The low-velocity mismatch between the real and simulated samples indicates that observations may currently be missing a large fraction of the population of outflowing \hi\ clouds (Yu et al., in prep.).
When comparing the six different epochs in the first panel of Fig. \ref{fig:clouds_comp_obs}, we find only a weak dependence of the velocity distribution on the level of star-formation activity, with more starbursting phases producing a population with a more extended high-velocity tail.

The line-of-sight velocity dispersion (Fig. \ref{fig:clouds_comp_obs}, second panel) spans values broadly consistent with the observations, ranging from about 8~$\kms$ (corresponding to the expected sound speed of a \hi\ gas at $T\sim10^4$ K) to a few tens of $\kms$.
The long high-velocity dispersion tail indicates that many clouds have a significant level of turbulence (see also Section~\ref{sec:clouds_general}). 
Here again, we note that the velocity dispersion increases, i.e.\ clouds tend to be more turbulent, during episodes of starbursts, as shown previously in Section~\ref{sec:clouds_time}. Our simulated clouds exhibit velocity dispersions that are typically about an order of magnitude lower than their bulk velocities, in agreement with the observed cloud population of \citet{DiT+18}. However, this does not appear to be a universal property of cool outflowing clouds. For example, \citet{2026arXiv260306568L} find that turbulence plays a major role in the kinematics of cool clouds in nearby star-forming galaxies, although their clouds are observed at larger distances from the disc ($\sim$10~kpc).

Observed clouds tend to have slighter larger velocity dispersions than their simulated counterparts, with a similar peak around $\sigma_\mathrm{v}\simeq15~\kms$ but extending to larger values.
This difference is particularly evident when comparing the data to simulation epochs characterised by low star formation activity. Furthermore, observed velocity dispersions also go to lower values than the $T\simeq10^4$~K gas sound speed (going as low as 5~$\kms$), which is not allowed for our simulated clouds.
Observational effects that broaden velocity profiles, such as beam smearing and instrumental broadening, may account for part of these discrepancies, but are unlikely to fully explain them.

Our simulation produces clouds with typical sizes in agreement with the observations (third panel), taking into account that the low radius end is artificially truncated for our simulated clouds at $\sim$12~pc to keep only resolved clouds (Section \ref{sec:Clouds}).
The simulated distributions of cloud radii at different times have similar shapes regardless of the underlying level of star formation in the CMZ, with the most frequent radius being $\sim15$~pc and with a tail extending to $50-60$ pc.
The observed distribution likewise peaks around $10-12$ pc and show a tail reaching up to $\sim40$ pc. 
We note that the large-radius end of the observed distribution may be affected by the source-finding criteria adopted to define the sample, which tend to exclude extended sources \citep[see][for details]{DiT+18}.
The apparent cut-off at small radii is also likely driven by resolution limits, affecting both the simulations and the observations. 

The \hi\ column densities of the simulated clouds (Fig. \ref{fig:clouds_comp_obs}, fourth panel) are consistent with the observed ones, with differences arising depending on the level of star formation in the CMZ, as starburst phases produce denser clouds (see Section~\ref{sec:clouds_time}). The observed cloud population peaks around $5 \times 10^{18}$~cm$^{-2}$, while the most frequent column densities in our simulation range from $0.3 \times 10^{18}$ to $10^{19}$~cm$^{-2}$ depending on the epoch considered. Interestingly, the post-starburst phase exhibits a tail extending to higher column densities ($\sim 2 \times 10^{20}$~cm$^{-2}$) than observed, whereas the third quiescent period produces substantially lower values ($\sim 10^{19}$~cm$^{-2}$).
Mass distributions in the simulated and observed cloud populations show broad agreement, although our simulation tends to under-produce the high-mass tail seen in the observations. In all cases, the bulk of the distribution lies around $100~\Msun$, but a few observed clouds reach masses of $10^4~\Msun$ and above, which is not reproduced in our simulations. Consistently with the column density distributions, the third quiescent period systematically produces lower cloud masses than both the observations and the other simulated epochs.

Overall, these comparisons highlight that our model of the MW nuclear region predicts the presence of SN-driven cool clouds whose average properties are broadly consistent with those observed in the GC.
Interestingly, even during periods of quiescent star formation activity in the simulation, \hi\ clouds are still present in the outflows, with number and properties roughly comparable to observations. 
This suggests that a strong, recent starburst in the CMZ is not required to account for the \hi\ cloud population observed at the GC. 
This conclusion is less robust for the third quiescent period, which produces fewer clouds with somewhat lower masses than observed. 
As discussed in Section \ref{sec:Outflows}, this phase corresponds to a time in which the SNER is extremely low, following an extended period of weaker star formation activity.
In this regime, the outflows generated in our simulation are indeed consistently weaker, leading to less clouds with lower masses (Fig. \ref{fig:clouds_sner}$a$-$e$).

Finally, we stress that, in the analysis presented above, we did not account for any of the observational effects that inevitably affect real data, such as the finite spatial and spectral resolution of the telescope or the sensitivity limits of the observations. 
A follow-up paper will present a more detailed and quantitative comparison with observations, explicitly incorporating these observational biases, and will attempt to predict cloud properties that are currently inaccessible in observations (e.g., the fraction of outflowing to inflowing clouds) from combinations of observable quantities. This approach will enable a more robust assessment of the similarities and differences identified above between the simulated and observed cloud populations.

\section{Discussion}
\label{discussion}

We have shown that our simulation forms a dense, CMZ-like ring in which the star formation and resulting outflows are regulated by the interplay between radial accretion and SN feedback.
The SN-driven multiphase outflows reach velocities of up to $\sim1000~\kms$ in the hot phase, while the cooler gas moves at a few hundred $\kms$.
Close to the disc, the outflows are dominated in mass by the cooler phases, whereas at larger heights the hot phase dominates both the mass and the energy budget. Cool clouds detected in the extraplanar region show properties consistent with entrainment and acceleration by the hot wind.
In the following, we place these results in the context of previous simulation studies of galactic outflows.
We consider three complementary classes of models: self-consistent simulations of CMZ-like structures, stratified-box simulations, and simulations focusing on the cold cloud production. In addition, we also discuss the main limitations of our current \textsc{Piernik} model.

\subsection{Comparison to CMZ-like models}
\label{comp_AREPO}
A number of works have simulated the nuclear regions of the MW \citep[e.g.][]{Rodriguez-Fernandez+2008,Kim+2011,Ridley+2017}, although with limited focus on the outflows. 
In particular, in recent years, several high-resolution models have used the moving-mesh code {\scshape Arepo} \citep{Springel2010} to investigate in detail the gas dynamics, morphology and star formation in the CMZ \citep{Sormani+17,Sormani+20,Tress+20,Tress+25}. 
The most advanced {\scshape Arepo} model to date, presented in \citet{Tress+25}, achieves a sub-pc resolution in the CMZ, but with a coarser resolution in the extra-planar region compared to our model (30 pc vs 3 pc). Their model uses the same external gravitational potential of our simulation, but includes more advanced physics, such as a chemical network, pre-stellar feedback and radiative transfer.
A detailed comparison between our model and that of \citet{Tress+25} is provided in Appendix~\ref{app:arepo}; here we summarise the main conclusions.

Both our and \citeauthor{Tress+25}'s simulations form a comparable $\sim$250 pc ring with similar surface density and vertical structure, but the evolution of their respective CMZ diverges. While our simulation sustains episodic accretion and starburst cycles (Section~\ref{sec:gas_dynamics}), in \citet{Tress+25}, the weakening of the bar-driven inflows lead to a declining SFR over time and a stabilisation into a quiescent CMZ, consistent with earlier {\scshape Arepo} models \citep{Tress+20,Sormani+20}.
This difference is reflected in the outflows: although \cite{Tress+25} launches hot winds, these are slower, less energetic, and largely unable to entrain cool gas beyond a few hundred pc, in contrast to the strong, multiphase outflows produced in our model. We attribute this mainly to differences in how SN feedback couples to the ISM: SNe in our simulation preferentially explode in lower-density gas and generate larger volumes of hot, fast material, whereas in \citet{Tress+25}, energy is injected mainly into denser gas where feedback is less disruptive.
This difference arises from a combination of the SN injection schemes and SN clustering. In our model, the fixed, volume-weighted injection scheme (Section~\ref{sec:FB}), together with stronger SN clustering due to lower resolution and the absence of pre-stellar feedback, favours explosions in low-density environments. In contrast, the mass-weighted, local injection and reduced clustering used in \citet{Tress+25} concentrate SN energy in denser gas.
While it remains unclear which simulation provides a more accurate description of the MW centre, this comparison highlights the sensitivity of the CMZ evolution and outflow properties to the adopted star formation and feedback prescriptions.

Additional recent simulations of MW-like CMZs have been presented in \citet{2019MNRAS.490.4401A} and \citet{Moon_2021}, exploring star formation and feedback.
In particular, \cite{2019MNRAS.490.4401A} found feedback-regulated SFR oscillations in their simulated CMZ, concluding that the present-day observed CMZ is at a minimum of such a cycle, and further investigated the corresponding outflows. Despite a much lower resolution outside the disc ($\sim 50$~pc at $n \sim 10^{-1}$~cm$^{-3}$ and $\sim 500$~pc at $n \sim 10^{-4}$~cm$^{-3}$), their outflow properties are overall similar to the ones produced in our simulation. They observe winds dominated by both a cool and a hot phase, the former falling back onto the disc in a fountain flow. While their mass loading factors are fairly consistent with ours, our stellar feedback seems to produce more hot gas, as evidenced by a higher contribution of the hot phase to the loading by a factor $\sim$1.5, and to much higher energy loadings (factor $\sim$3) heavily dominated by the hot gas (see Section \ref{sec:Outflows}).
\cite{Moon_2021} studied star formation in nuclear rings, with their L2 model showing ring surface density and SFR density comparable to those of our simulated CMZ. Consistent with our findings, the outflows produced by this ring reach velocities of nearly 1000~$\kms$ in the hot phase, and have a total mass loading similar to ours.

\subsection{Comparison to multiphase wind models}

Several high-resolution simulations of multiphase outflows have been performed using stratified (or ``tall'') box setups, which model only a small, vertically-extended patch of a galactic disc \citep[e.g.][]{2017ApJ...841..101L,2020ApJ...900...61K,2020ApJ...890L..30L,2023MNRAS.522.1843R,2025MNRAS.539.1706V}. Direct comparison with such models is not straightforward because none of them explicitly models the nuclear rings of a barred galaxy, but rather portions of discs with similar gas and star formation surface densities. 
Therefore we do not expect a strict quantitative agreement.
Models with conditions similar to our CMZ 
generally find that the mass loading remains below unity and the energy loading is systematically dominated by the hot gas. However, the relative contributions of the hot and cool phases to the mass loading and the overall energy loading factor vary between studies. Some models, similarly to ours, produce powerful hot winds that dominate the mass loading while contributing energy loading factors of order a few tenths \citep[e.g.][]{2017ApJ...841..101L,2025MNRAS.539.1706V}. In contrast, others studies find that the cool phase dominates the mass loading, whereas the hot phase results in comparatively lower energy loading \citep[e.g.][]{2020ApJ...900...61K,2023MNRAS.522.1843R}.
For all models, the velocities of the hot wind commonly reach values close to $1000~\kms$ while the cool phase is slower with velocities around $100~\kms$ \citep[e.g.][]{2017ApJ...841..101L,2020ApJ...900...61K,2023MNRAS.522.1843R}, consistently with our simulated outflows. As shown, for example, by \cite{2020ApJ...890L..30L}, the hot gas loading factors depends on whether a substantial fraction of SN events occurs in low-density gas. In such cases, a significant fraction of SN mass and energy can escape into the halo. This situation can occur, for instance, if the SN scale height exceeds the gas scale height, so that SNe explode at high latitudes where the ambient density is low, as in  \cite{2017ApJ...841..101L} and \cite{2025MNRAS.539.1706V}, where the SN scale height is imposed by hand. In our model, although the SN locations are naturally set by star formation, a large fraction of SNe still explode in low-density gas. This is mainly because most of the gas is concentrated in a thin ring, leaving the rest of the volume filled by low-density material. Stars formed in the dense ring can migrate away from their birth sites, and some eventually explode in more diffuse regions.

A few recent works focussed on studying the properties of cool gas outflowing from the nuclear regions of star-forming galaxies. These properties depend on the star-forming environment generating the wind. Most simulations to date have been tuned to reproduce the conditions of M82 \citep[e.g.,][]{Schneider+18,Schneider+20,Tan+24,2025ApJ...984..191W}, the most studied starburst galaxy in the nearby universe. Cool gas clouds in a MW-like environment have been studied only recently in \citet{2025ApJ...981..154Z}, albeit for a system in a significantly more starbursting phase than the present-day CMZ (i.e.\ ${\rm SFR}\sim 2~\msunyr$).
The presence of a higher SFR combined with a shallower gravitational potential in M82-like environments drives more powerful winds able to accelerate clouds to very large velocities, reaching several hundred to a thousand $\kms$ \citep{Schneider+20,2025ApJ...984..191W} at $|z|\sim2$~kpc, while MW-like winds produce a gentler acceleration, with velocities limited to a few hundred $\kms$, comparable to those in our simulation. As in our work, clouds are consistently found to be thermally under-pressured relative to the surrounding hot medium \citep[e.g.,][]{Schneider+20}; however, their significant turbulent pressure can compensate for this deficit, bringing them close to pressure equilibrium with the hot phase \citep{Tan+24,2025ApJ...984..191W}. This is also consistent with \citet{10.1093/mnras/staf949}, who demonstrate the importance of dynamical pressure for cold clouds embedded in a hot wind, as it helps support the clouds against compression induced by radiative cooling.
Compared to our findings, clouds in a M82-environment \citep[e.g.,][]{Tan+24,2025ApJ...984..191W} show similar ranges of masses, densities, temperatures, volumes and density contrasts.
Despite their larger outflow speeds, cloud velocity dispersions are also comparable to our values ($\sigma_\mathrm{v} \sim 20~\kms$).
As in our simulation, cloud velocities, densities, and sizes increase with distance from the galactic mid-plane, while their masses decrease \citep{2025ApJ...984..191W, 2025ApJ...981..154Z}.
In agreement with our conclusions in Section~\ref{sec:clouds_survival}, these studies generally find the \cite{Gronke+19} criterion to be overly permissive for cloud survival \citep[e.g.,][]{Schneider+20}, and instead prefer the criteria proposed by \cite{Abruzzo+22} or \cite{Farber&Gronke+22} \citep[e.g.,][]{Tan+24,2025ApJ...984..191W}. Accordingly, \cite{2025ApJ...981..154Z} report that only $\sim$5\% of clouds survive in a MW-like GC environment.

Finally, a recent work by \citet{Afruni+26}, used a semi-analytic approach to explore the launching and acceleration of cold clouds by a hot, SN-driven wind in the GC. To reproduce the properties of \hi\ clouds observed in the GC, they require a hot wind with a mass loading factor $\eta_\mathrm{hot}\sim0.1$, an energy loading factor $\beta_\mathrm{hot}\sim0.4$, and velocities of order 1000~$\kms$. These values are broadly consistent with the hot phase in our simulation (see Fig. \ref{fig:zprofs_piernik}). As in our model, such a wind leads to positions and velocities of cold clouds comparable to observations, supporting the interpretation that the high-velocity \hi\ clouds observed above the GC can be explained by SN-driven hot winds.

\subsection{Limitations of our model}
\label{sec:limitations}

Below, we discuss the most important caveats of our simulation and their potential impact on our results.

As noted in the comparison with the simulation by \cite{Tress+25}, our model does not include pre-supernova feedback, such as radiation pressure or stellar winds, which can disperse gas in molecular clouds and thereby reduce SN clustering. Additionally, sampling an initial mass function would introduce stochastic variations in SN timing and multiplicity. Our model does neither: all star particles are identical and explode after a fixed lifetime of 6.5 Myr. This delay is only an approximation, as the time between star formation and a SN event is set by stellar evolution and depends on progenitor mass. More massive stars evolve more rapidly and explode as SNe on shorter timescales than their lower-mass counterparts, implying that in reality some SNe are expected to occur earlier than 6.5~Myr, and others later, up to a few tens of Myr. This simplification is expected to affect both SN clustering (earlier explosions can disperse the dense gas in which stars form, suppressing local star formation and subsequent SNe) and the ambient density at which SNe explode, as longer delays allow star particles to migrate away from their birth environment. As discussed in Section~\ref{comp_AREPO}, both effects can potentially influence the efficiency of feedback, and consequently the resulting outflow production.

Our simulation achieves a spatial resolution of 3~pc in the CMZ and in the dense gas above it. To assess the impact of resolution, we performed test runs with reduced resolution in the outflowing region, keeping the CMZ resolution unchanged. The global wind properties, such as the mass and energy outflow rates, remain largely unchanged across all gas phases, except for the cold component ($T<5000$ K). This likely reflects the fact that outflows are launched from the disc, where the resolution is identical, so the various phases persist to large heights regardless of resolution, with the exception of dense cold gas, which is characterized by smaller scales compared to the other phases and therefore requires higher resolution to avoid excessive mixing with the hot wind. 
We therefore conclude that a resolution of 3~pc is sufficient to capture the global multiphase structure of the ISM and the large-scale dynamics of the outflows.

Clearly, individual cloud properties are more sensitive to the adopted resolution. For this reason, we chose to focus on the statistical properties of the cloud population rather than on a detailed analysis of individual clouds. Indeed, a reduced resolution is likely to increase numerical diffusion and mixing with the surrounding hot phase, leading to more rapid cloud evaporation into the surroundings. As mentioned in Section \ref{sec:Clouds}, previous works have shown that convergence in individual cloud properties requires the cloud radius to be resolved by at least four resolution elements for subsonic clouds and eight resolution elements for supersonic clouds. Therefore, since most of our simulated clouds are subsonic, we limited our analysis to clouds with more than 4 cells per cloud radius. Also, we notice that the adopted resolution affects the low-mass tail of the cloud population, since the minimum cloud radius is effectively set by the resolution scale.

Finally, our model does not include a central supermassive black hole. The presence of an AGN in the MW's history remains uncertain, as Sgr A$^*$ is currently in a relatively quiescent state \citep[e.g.][]{2014ARA&A..52..529Y}. Accordingly, the work presented here focuses exclusively on outflows driven by SN feedback, without AGN contributions. Nevertheless, Sgr A$^*$ may have experienced periods of enhanced activity in the recent past \citep[e.g.,][]{2012ApJ...756..181G,Bland-Hawthorn+2019,2020ApJ...894..117Z}, which might have contributed to the present-day outflows observed in the nuclear region. While modelling AGN feedback is beyond the scope of this work, its addition would likely drive stronger outflows, potentially accelerating a larger number of cool clouds to higher velocities.

\section{Summary and conclusions} 
 \label{ccl}
Using the {\scshape Piernik} code, we ran a simulation of the Milky Way, focusing on the Central Molecular Zone (CMZ) and the gaseous winds launched from it. Our simulation includes magneto-hydrodynamics, self-consistent supernova feedback, and thermal gas reaching 3~pc resolution in the CMZ and in the nuclear outflows. We summarise our results below: 

\let\labelitemi\labelitemii
\begin{itemize}[label=$\bullet$]
    \item The star formation rate in the CMZ varies significantly over time, ranging from $\sim$0.1~$\msunyr$ to $\sim$1~$\msunyr$ and showing cycles of starbursts followed by quiescent periods. These star-formation cycles are regulated by a combination of radial inflows from the bar's dust lanes, feeding the CMZ with fresh gas, and stellar feedback, which removes gas through outflows and inhibits gas accretion.

    \item Supernova feedback alone can generate significant multiphase outflows in the extraplanar region. Outflows are energetically dominated by the hot gas phase ($T>10^6$~K), which accelerates colder gas ($T<2\times10^4$~K) up to 2~kpc from the mid-plane. This colder gas dominates the wind mass budget at heights $|z|<300$~kpc. Periods of starbursts lead to stronger outflows from all gas phases, while quiet star formation periods are less efficient at creating winds.
    However, due to the time offset between SFR and supernovae, significant outflows can still occur during phases of relatively-low SFR, following a starburst event.

    \item Numerous extraplanar cool gas ($T\sim10^4$~K) clouds, tracing the atomic hydrogen (\hi) phase of the wind, are found in the simulation. These clouds are either outflowing from or inflowing onto the disc with velocities of the order of a few hundred $\kms$. The properties of the simulated clouds are in broad agreement with the \hi\ population observed above and below the Galactic centre, in terms of masses, position, velocities, velocity dispersions and radii. 

    \item Cool clouds in the outflows are entrained and accelerated by the hot wind. As they move away from the disc mid-plane, clouds mix with and lose mass to the surrounding hot gas, becoming increasingly elongated along the wind direction, warmer, and more diffuse.

    \item The median properties of simulated \hi\ clouds change over time, reflecting the evolution of the underlying star-forming environment: more supernovae lead to a larger population of outflowing clouds, with larger masses and densities, lower temperatures, higher velocities and velocity dispersions.

    \item About $20\%$ of extraplanar clouds are falling back onto the disc, indicating the presence of fountain flows above the CMZ. Infalling clouds are found closer to the disc, and are typically less massive, larger and less dense than outflowing ones at similar heights. The small fraction of inflowing clouds, together with the small amount of clouds reaching $|z|=2$~kpc, suggests that most outflowing clouds do not survive and are destroyed by the hot wind, and excludes the \citep{Gronke+19} criterion for survival in our simulation as being overly optimistic for our simulated clouds.

\end{itemize}

\noindent In conclusion, our simulation provides a self-consistent model of the inner Milky Way that connects gas inflow into the CMZ, cyclic star formation, and supernova-driven feedback to the formation of multiphase nuclear outflows.
A key result of this work is that supernova feedback alone, acting within a time-variable star-forming environment, is sufficient to generate hot, energetic winds capable of entraining and accelerating cool gas clouds to kiloparsec heights. 
Although quiescent periods produce lower-intensity outflows, the latter remain powerful enough to produce outflowing clouds consistent with observations.
This could be a natural explanation for the presence of high-velocity \hi\ clouds observed above and below the Galactic centre, despite the presently modest level of star formation in the CMZ.

In future work, we will extend this framework by including additional physical processes relevant for driving outflows (e.g. cosmic rays and AGN feedback) and by confronting the simulated winds with multi-wavelength observational tracers of outflows in the Galactic centre, including Fermi and eROSITA bubbles, UV absorption, and \hi\ clouds.

\vspace{0.2cm}
\begin{acknowledgements}
The authors would like to thank Mattia C.\ Sormani for very helpful discussions on the gravitational potential building up the CMZ, and the anonymous referee for constructive feedback and comments.
EDT and NP were supported by the European Research Council (ERC) under grant agreement no.\ 101040751.
LA acknowledges support through the Program ``Rita Levi Montalcini'' of the Italian MIUR. 
Calculations were carried out at the centre of Informatics - Tricity Academic Supercomputer \& networK (TASK), on the HYDRA cluster at the Institute of Astronomy of Nicolaus Copernicus University in Toruń.
We further acknowledge ISCRA for awarding this project access to the LEONARDO supercomputer, owned by the EuroHPC Joint Undertaking, hosted by CINECA (Italy).
\end{acknowledgements}

\bibliographystyle{aa} 
\bibliography{biblio}

\begin{appendix}

\section{Comparison with \cite{Tress+25}}
\label{app:arepo}

In this Appendix, we present a detailed comparison between our model and the simulation of the GC by \cite{Tress+25} (\citetalias{Tress+25}, hereinafter), briefly discussed in Section~\ref{comp_AREPO}. Direct access to the output data of this model allowed us to perform an independent analysis of the outflows generated by stellar feedback in their CMZ. This simulation is presented in detail in \citetalias{Tress+25}, but we provide below a brief summary of its most relevant features.  

\subsection{Presentation of \cite{Tress+25} model}
\label{app:AREPO_presentation}

The model uses the moving-mesh code {\scshape Arepo} \citep{Springel2010,Weinberger2020} to simulate a MW-like galaxy.
The authors adopt the same potential as in our model \citep{2024A&A...692A.216H}, although including the contributions of the central supermassive black hole and of the spiral arms, which were neglected in our simulation. The whole Galactic disc is modelled up to radius $R=10$~kpc, with enhanced resolution in two particular regions: the solar circle ($7.5<R<8.5$~kpc) and the Galactic centre ($R<4$~kpc), where the mass resolution reaches 20~$\Msun$ compared to 500~$\Msun$ in the rest of the disc and in the halo. 
Within 1~kpc distance from the mid-plane, cells also have a maximum volume of 10$^5$~pc$^3$ ($r_\mathrm {cell}\sim30$~pc). Furthermore, the simulation also includes a central sink particle representing the supermassive black hole Sgr A$^*$. 
The first 100~Myr of the simulation include only the axisymmetric components of the gravitational potential, at lower resolution, to establish an approximate equilibrium state. 
Afterwards, the non-axisymmetric components are gradually introduced over 150~Myr (similarly to our model) to form the Galactic bar and the CMZ.

Chemistry of hydrogen and CO gas is followed thanks to a time-dependent chemical network \citep{Glover2012}. 
This provides self-consistent radiative heating and cooling for temperatures below 10$^4$~K, while for higher temperatures they are determined by tabulated values. 
Star formation is modelled with a numerical scheme where star particles are stochastically created with an efficiency of 10\% from gas particles at a local minimum of the gravitational potential, with negative velocity divergence \citep{Goller2025}. 
An initial mass function sampling is then used to populate star particles with O-B stars on a sub-grid level, which will generate ionising radiation feedback for the duration associated to their main-sequence lifetime. The radiative transfer equations are then solved using Sweep \citep{Peter2023} in {\scshape Arepo}. 
Supernova feedback is implemented both as type Ia (linked to old stellar populations, distributed randomly), and type II, which are coupled to star particles and explode at the end of the main sequence lifetime of their stellar population. 
Supernova events inject 10$^{51}$ erg of energy into the nearest neighbour cells, either in the form of thermal energy if the adiabatic Sedov-Taylor phase is resolved ($\sim$90\% of  SNe), or kinetic energy if it is not.  

The main advantage of this simulation over our model is that it includes more advanced physics such as a chemical network, pre-stellar feedback, and radiative transfer. It also achieves higher resolution in the dense gas, reaching sub-pc scales (down to a minimum of 0.2~pc), compared to the 3~pc resolution in the CMZ of our model. However, the resolution in \citetalias{Tress+25} is lower in the outflows, of the order of 30~pc above the CMZ, while in our model we maintain 3~pc resolution in the denser outflowing gas. Likewise, in the low-density gas of the CMZ, the lowest resolution in our simulation is 6~pc, compared to the 30~pc of \citetalias{Tress+25}.

\subsubsection{Results of \cite{Tress+25} model}
\label{sec:AREPO_results}

\begin{figure}
\includegraphics[width=0.51\textwidth,trim = 0.8cm 0 0 2cm, clip]{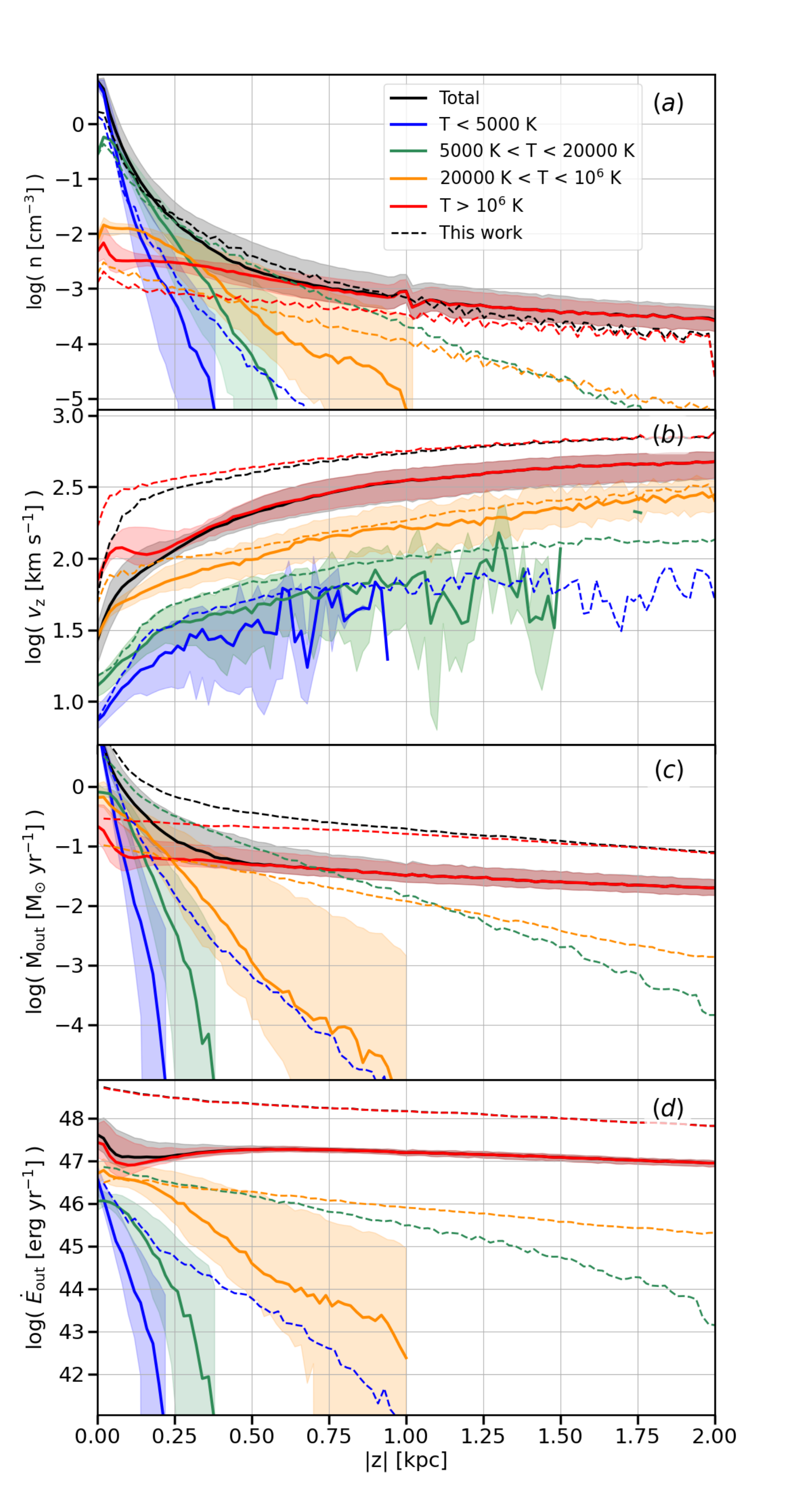}
\caption{Same as Fig. \ref{fig:zprofs_piernik}, but for the \citetalias{Tress+25} model (solid lines) compared to our model (dashed lines), and with the panel (c) and (d) showing the mass and energy outflow rates instead of the loading factors.}
    \label{fig:zprofs_arepo}
\end{figure}

   \begin{figure}
    \includegraphics[scale=0.55]{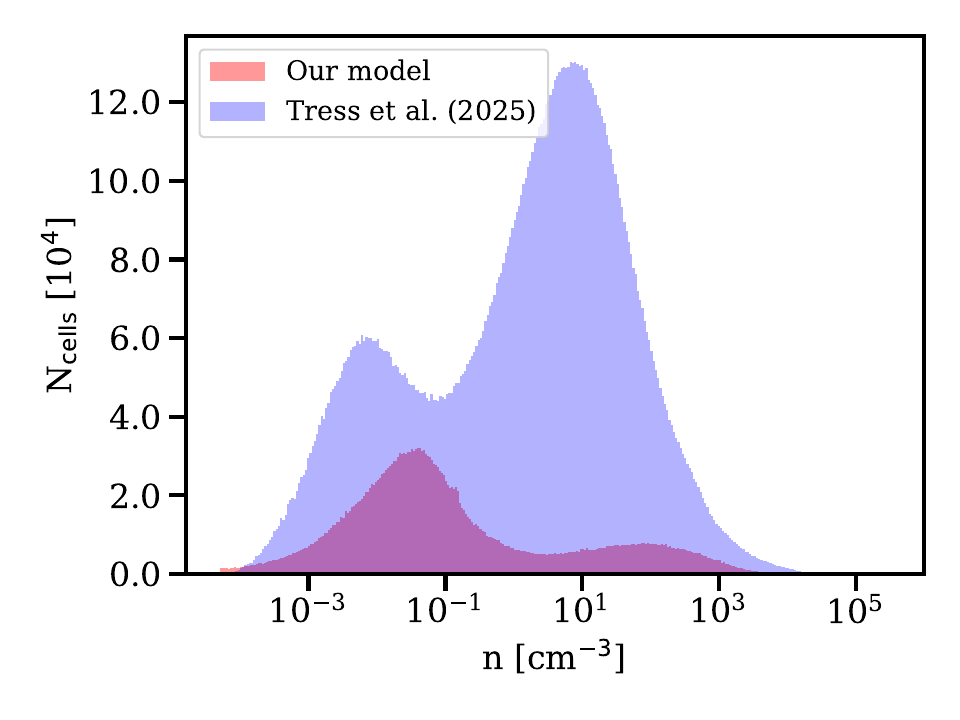}
    \caption{Distribution of gas cell densities where supernova energy is injected throughout the whole simulation, for our {\scshape Piernik} and the \citetalias{Tress+25} models. Supernovae tend to explode in lower density cells in our simulation than in \citetalias{Tress+25}, therefore creating more hot gas and winds. \citetalias{Tress+25} gas cells are typically smaller than in our model, explaining why the cells counts are higher for the former. }
    \label{fig:SNdens}
   \end{figure}

\citetalias{Tress+25} presents a detailed analysis of the morphology and dynamics of the CMZ in their model, with an emphasis on the star formation regulation, and the decoupling between stars and gas. We therefore refer the reader to this paper for more details on these aspects, and focus here on the comparison to our model, with particular emphasis on the outflow properties.

\citetalias{Tress+25}'s simulation forms a clear CMZ ring of $\sim$250~pc radius, whose surface densities and extent are comparable to those in our model. The two discs seen edge-on also show excellent agreement, being of similar thickness and temperature, which translates to comparable vertical density profiles for different temperature phases within the disc. However, over time, in the \citetalias{Tress+25} simulation the dust lanes on each side of the ring become weaker, while the CMZ becomes smoother and more regular, which is not something we observe in our model. 

In the \citetalias{Tress+25} model, while initially the SFR is very high ($>1~\msunyr$), it quickly decreases to reach lower values between 0.05 and 0.3~$\msunyr$, and stays low until the end of the simulation. The SNER follows the SFR, and also decreases over time towards $\sim$10$^{48}$~erg~yr$^{-1}$. It therefore appears that the CMZ stabilises over time, reaching a rather quiet state with a SFR comparable to the present-day CMZ. This is linked to a global decline of the radial inflows feeding the CMZ through the dust lanes, as the latter gradually weaken. With little gas brought to the CMZ, the gas mass available in the CMZ decreases over time due to gas removal from star formation and outflows, causing in turn the SFR to decline as well. The episodic radial accretion present in our model leading to starburst cycles (see Section \ref{sec:CMZ}) is therefore not observed in their model.
Following the global trend of the SFR, the mass outflow rate $\dot{M}_\mathrm{out}$ in \citetalias{Tress+25} decreases over time: while initially similar to ours ($\sim0.3~\msunyr$ at $|z|=300$~pc), it quickly drops to lower values (around 0.05~$\msunyr$ at $|z|=300$~pc, similar to the very end of our simulation).

In Fig.~\ref{fig:zprofs_arepo}, we compare vertical profiles of gas density (panel $a$), vertical velocity ($b$), mass outflow rate ($c$) and energy outflow rate ($d$) between our simulation (dashed lines) and that of \citetalias{Tress+25} (full lines).
The total $\dot{M}_\mathrm{out}$ averaged over time is about a factor of 6 lower than in our model at $|z|\simeq300$~pc, and is dominated by the hot phase (Fig. \ref{fig:zprofs_arepo}$c$). The differences become even more pronounced at larger heights: in their model, $\dot{M}_\mathrm{out}$ for all phases below $10^6$~K declines much more steeply with $|z|$ than in our simulation, leading to significant discrepancies in the vertical profiles of both density and mass outflow rate between the two models (panels $a$ and $c$, respectively). In \citetalias{Tress+25}, outflows are consequently largely dominated in mass by the hot gas outside the disc (for $|z|>$1~kpc it is even the only phase remaining), and totally devoid of cool material ($T<2\times10^4$~K) for $|z|>$500~pc.

The energy outflow rates show even larger differences (Fig.~\ref{fig:zprofs_arepo}$d$): the \citetalias{Tress+25} model exhibits total $\dot{E}_\mathrm{outf}$ that are about an order of magnitude lower than those in our model at all heights, due to the lower $\dot{E}_\mathrm{outf,hot}$ of the hot phase. Consistently, the volume filling fraction of hot gas in the ISM is lower in \citetalias{Tress+25} than in our simulation, which indicates a less efficient production of hot gas by SN feedback. The resulting vertical velocities of the hot outflowing gas (Fig. \ref{fig:zprofs_arepo}$b$) are, as expected, lower in \citetalias{Tress+25} than in our model by a factor of $\sim$2: they never exceed $\sim800~\kms$ in \citetalias{Tress+25}, while in our simulation they can reach up to $\sim 5000~\kms$. This leads to lower velocities of the colder phases as well, with the cool gas ($5000<T<2\times10^4$~K) in \citetalias{Tress+25} accelerated to only a few tens of $\kms$ and typically only within a few hundred pc (panel $c$ shows that $\dot{M}_\mathrm{outf,cool}$ becomes negligible for $|z|\gtrsim300~$pc).
Since \citetalias{Tress+25} includes a chemical network with access to \hi, H$_2$, He, H$^+$ and CO gas density, we analysed the presence of these different elements in the outflows. Consistently with the temperature phases analysed above, we found that while hydrogen in the disc is predominantly neutral, the outflows are dominated by ionized gas (H$^+$), with small amounts of \hi\ up to $|z|=500$~pc, and no molecular gas outside the disc.
\\

It therefore appears that, although stellar feedback is capable of launching hot winds from the CMZ in the \citetalias{Tress+25} model, these winds are significantly slower and less energetic than in our simulation. Their reduced strength naturally explains the near absence of cool \hi\ (and molecular) gas entrainment in the outflows of \citetalias{Tress+25}, in contrast to our findings.
While it is beyond the scope of this paper to make a detailed comparison of the {\scshape Arepo} and {\scshape Piernik} codes, we attempt below to explore what could be responsible for this discrepancy. 

Supernova feedback is more disruptive in our model: unlike in \citetalias{Tress+25}, it can break up the dust lanes, leading to episodic radial accretion onto the CMZ while generating very fast hot outflows. We tentatively explain this discrepancy with Fig.~\ref{fig:SNdens}, which shows the number density distributions of the gas cells in which SN energy is deposited throughout our simulation (red) and that of \citetalias{Tress+25} (blue). Each SN injects energy into several gas cells (as described in Section~\ref{sec:FB} and Appendix~\ref{app:AREPO_presentation}), so that the gas cell count shown in Fig. \ref{fig:SNdens} is higher than the number of SNe exploding in the entire simulation. Interestingly, we observe a bimodal distribution of the density in both simulations, with a low density peak around  10$^{-2}$ cm$^{-3}$ (consistent with previous studies, e.g. \citealt{2020ApJ...891....2L} for type II SNe in young stellar clusters), and a second higher density peak around $10-100$~cm$^{-3}$. The main difference between the two models however, is that while most of the SN energy in our simulation is deposited at low densities, it is injected into high density cells in \citetalias{Tress+25}. This explains at least partially why our model produces hotter, higher velocity gas: SNe in high density gas are less disruptive and produce less hot gas. 

This discrepancy in the SN injection cells is likely due to a combination of two factors: the SN injection scheme and clustering. While in \citetalias{Tress+25} the SN energy is injected only to neighbouring cells and is weighted by mass (explaining the peak of injection for small, high density cells in Fig. \ref{fig:SNdens}), in our model the fixed injection area (3 cell radius) with a volume-weighted injection rate means low density cells generally receive more energy. 
Moreover, using the Ripley's $L$ function \citep{10.1111/j.2517-6161.1977.tb01616.x} measuring whether data points have a random, dispersed or clustered distribution, we found that SNe in our simulation tend to have a higher clustering coefficient than in \citetalias{Tress+25}. Clustered SNe are known to produce more disruptive feedback and stronger winds \citep[e.g.][]{2017MNRAS.470L..39F}, as they clear out the region surrounding them, making them more likely to explode in low density environments. The higher clustering present in our model could be due to the lower resolution in high density gas and lack of pre-stellar feedback, leading to larger star forming clumps \citep{2021MNRAS.506.3882S}. This may explain why the pre-stellar feedback in \citetalias{Tress+25} creates less clustered SNe that inject energy into a denser medium, rather than reducing the ambient density at SN sites.
  
These differences in the ISM physics and in the implementation of SN feedback likely account for why the \citetalias{Tress+25} model produces a quieter GC, with a more regular, steady CMZ and a stable SFR around 0.1~$\Msun$~yr$^{-1}$, matching that of the present-day CMZ, but with very little outflows. In contrast, our model produced a more turbulent CMZ characterized by episodic gas accretions and starburst phases that drive strong outflows.
Determining which model more accurately reproduces the GC conditions is not straightforward. While the \citetalias{Tress+25} model incorporates more sophisticated physics, it fails to produce outflows powerful enough to match observations in the GC using star formation feedback alone. Based on this model, one might conclude that additional mechanisms, such as an AGN or cosmic rays \citep[e.g.][]{1993A&A...269...54B, 2013ApJ...777L..38H,2018MNRAS.479.3042G}, are required to account for the cold clouds observed in the MW nuclear outflows.
On the other hand, our model suggests that such additional processes may not be necessary.

\end{appendix}

\end{document}